\documentclass[eprint,twocolumn,aps,prx,longbibliography,amsmath,amssymb]{revtex4-1}
\usepackage[latin9]{inputenc}
\usepackage{mathrsfs}
\usepackage{amstext}
\usepackage{amssymb}
\usepackage{graphicx}
\usepackage{esint}
\usepackage[export]{adjustbox}[2011/08/13]
\usepackage{hyperref}
\usepackage{pgfplots}
\pgfplotsset{compat=newest}
\usepgfplotslibrary{groupplots}
\usepgfplotslibrary{dateplot}

\makeatletter
\usepackage{epsfig}
\usepackage{bm}
\usepackage{dcolumn}
\usepackage{color}

\newcommand{\scS}{\mathcal{S}}

\newcommand{\fourvec}[4]{\left(\begin{array}{cc} #1 & #2 \\ #3 & #4 \end{array} \right)}

\makeatother

\begin{document}

\title{Beyond Gaussian pair fluctuation theory for strongly interacting Fermi gases II: The broken-symmetry phase}

\author{ Brendan C. Mulkerin$^{1,2,7}$, Xing-Can Yao$^{3,4,5}$, Yoji Ohashi$^{6}$, Xia-Ji Liu$^{7}$ and Hui Hu$^{7}$} 

\affiliation{$^{1}$School of Physics and Astronomy, Monash University, Victoria 3800, Australia}

\affiliation{$^{2}$ARC Centre of Excellence in Future Low-Energy Electronics Technologies, Monash University, Victoria 3800, Australia}

\affiliation{$^{3}$Hefei National Laboratory for Physical Sciences at the Microscale and Department of Modern Physics, University of Science and Technology of China, Hefei 230026, China}

\affiliation{$^{4}$Shanghai Branch, CAS Center for Excellence in Quantum Information and Quantum Physics, University of Science and Technology of China, Shanghai 201315, China}

\affiliation{$^{5}$Shanghai Research Center for Quantum Sciences, Shanghai 201315, China}

\affiliation{$^{6}$Department of Physics, Keio University, Hiyoshi, Kohoku-ku, Yokohama 223-8522, Japan}

\affiliation{$^{7}$Centre for Quantum Technology Theory, Swinburne University of Technology, Melbourne, Victoria 3122, Australia}

\date{\today}

\begin{abstract}
We theoretically study the thermodynamic properties of a strongly interacting Fermi gas at the crossover from a Bardeen-Cooper-Schrieffer (BCS) superfluid to a Bose-Einstein condensate (BEC), by applying a recently outlined strong-coupling theory that includes pair fluctuations beyond the commonly-used many-body $T$-matrix or ladder approximation at the Gaussian level. The beyond Gaussian pair fluctuation (GPF) theory always respects the exact thermodynamic relations and recovers the Bogoliubov theory of molecules in the BEC limit with a nearly correct molecule-molecule scattering length. We show that the beyond-GPF theory predicts quantitatively accurate ground-state properties at the BEC-BCS crossover, in good agreement with the recent measurement by Horikoshi \textit{et al.} in Phys. Rev. X \textbf{7}, 041004 (2017). In the unitary limit with infinitely large $s$-wave scattering length, the beyond-GPF theory predicts a reliable universal energy equation of state up to 0.6$T_c$, where $T_c$ is the superfluid transition temperature at unitarity. The theory predicts a Bertsch parameter $\xi \simeq 0.365$ at zero temperature, in good agreement with the latest quantum Monte Carlo result $\xi = 0.367(7)$ and the latest experimental measurement  $\xi = 0.367(9)$. We attribute the excellent and wide applicability of the beyond-GPF theory in the broken-symmetry phase to the reasonable re-summation of Feynman diagrams following a dimensional $\epsilon$-expansion  analysis near four dimensions ($d=4-\epsilon$), which gives rise to accurate predictions at the second order $\mathcal{O}(\epsilon^2)$. Our work indicates the possibility of further improving the strong-coupling theory of strongly interacting fermions based on the systematic inclusion of large-loop Feynman diagrams at higher orders $\mathcal{O}(\epsilon^n)$ with $n\ge 3$.
\end{abstract}

\maketitle

\section{Introduction}  
There has been considerable effort over the last two decades to explore the properties of ultracold strongly interacting atomic Fermi gases \cite{Giorgini2008,Bloch2008,Randeria2014}.  In particular, the rapid advancement of experimental techniques has allowed for increasingly accurate measurements \cite{Nascimbene2010,Horikoshi2010,Ku2012,Horikoshi2017}  of the universal thermodynamics of superfluid Fermi gases \cite{Ho2004,Hu2007}. 
This progress gives us access to understanding other less accessible or controllable strongly interacting Fermi systems, such as  nuclear matter \cite{Lee2006,Hen2014,vanWyk2018,Ohashi2020} and high-$T_c$ superconductors \cite{loktev2001,Stajic2017}.

In greater detail, with the use of Feshbach resonances \cite{Chin2010} the interactions between fermions with unlike spins can be tuned to create weakly-coupled Cooper pairs and tightly-bound bosonic molecules, realizing a smooth crossover from a Bardeen-Cooper-Schrieffer (BCS) Fermi superfluid to a Bose-Einstein condensate (BEC) of bosons at sufficiently low temperatures \cite{Randeria2014}. Near the Feshbach resonance, the $s$-wave scattering length $a_s$ becomes infinite and the system is uniquely defined by a single energy scale, the Fermi energy $E_{\rm F}$, or by the corresponding temperature scale $T_{\rm F}=E_{\rm F}/k_B$. In this unitary regime the system is scale invariant and its thermodynamic properties appear to be universal \cite{Ho2004}: all the thermodynamic functions are a function of a single parameter $T/T_{\rm F}$, independent of the microscopic details of the underlying interactions. As pointed out in Ref.~\cite{Nikolic2007}, the universality of ultracold Fermi gases extends beyond the unitary regime to the BEC-BCS crossover. Interestingly, the general universality might be characterized by some universal relations, first derived by Shina Tan \cite{Tan1,Tan2,Tan3}. These relations reveal that the short-range and large-momentum behavior of dilute Fermi gases is encapsulated in the so-called contact parameter, which connects the thermodynamic properties of a fermionic system to its short-range behavior \cite{Braaten2008,Zhang2009,Stewart2010,Kuhnle2010,Werner2012,Braaten2012}. Due to the universality, the accurate determination of the thermodynamic properties of ultracold atomic Fermi gases would provide insight into other strongly interacting Fermi systems. 

From a theoretical perspective, understanding ultracold strongly interacting fermions at finite temperatures is a grand challenge \cite{Randeria2014}. As there is no small interaction parameter to perturbatively expand about,  approximate perturbative methods in general are uncontrollable. Therefore, sophisticated quantum Monte Carlo (QMC) techniques are often used in the theoretical studies of strongly interacting Fermi systems. QMC calculations have been successfully applied to zero temperature superfluids \cite{Forbes2011,Carlson2011,Drut2012}, finite temperature superfluids \cite{Goulko2016QMC,Jensen2020,Halford2020}, and for systems above the superfluid transition in the normal state \cite{VanHoucke2012,Drut2012b,Rossi2018}. These QMC simulations are varied in their methodology, and with the difficulty in finding the zero-effective-range equation of state, the converged results differ between schemes. The most accurate simulation achieved so far is the calculation of the normal-state density or pressure equation of state of the unitary Fermi gas, through the bold diagrammatic Monte Carlo (BDMC) technique \cite{VanHoucke2012}. This state-of-the-art theoretical result has a relative accuracy at the level of a few percent, comparable to the precision of the experimental data \cite{Ku2012,VanHoucke2012}. However, the application of the intriguing BDMC approach to a superfluid Fermi gas is yet to be demonstrated.

Field theoretical methods offer an alternative route to calculate the thermodynamic properties without the difficulty of convergence. These include various diagrammatic many-body $T$-matrix theories within the ladder approximation \cite{nozieres1985bose,de1993crossover,Ohashi2003,Pieri2004,Chen2005,Liu2005PRA,Taylor2006,Hu2006,Haussmann2007,Diener2008,Huicomparitive08,Watnabe2010,He2015,Mulkerin2016,Tajima2017,Mulkerin2017,Pini2019}, the dimensional $\epsilon$-expansion \cite{Nishida2006,Arnold2007,Son2007,Nishida_2007}, large-$N$ expansion \cite{Veillette2006,Nikolic2007}, and non-perturbative functional renormalization group approach \cite{Diehl2007,Boettcher2014}. 
Focusing on the different many-body $T$-matrix theories, where the complete geometric series of ladder diagrams can be summed up. The difference lies on the degree of self-consistency taken in the diagrams, since either the bare non-interacting Green function or dressed interacting Green function can be used. In the normal state, the advantages of different $T$-matrix theories have been comparatively reviewed in Ref. \cite{Huicomparitive08} (see also the recent extended study in Ref. \cite{Pini2019}). In this work, we are interested in the broken-symmetry superfluid phase below the critical temperature $T_c$. 

As far as ultracold superfluid Fermi gases are concerned, the non-self-consistent $T$-matrix theory was first formulated by Ohashi and Griffin \cite{Ohashi2003} and by Pieri, Pisani and Strinati \cite{Pieri2004}. A slightly modified version, in light of the seminal work by Nozi\`{e}res and Schmitt-Rink \cite{nozieres1985bose}, was also developed shortly after \cite{Hu2006}. The modification is equivalent to considering pair fluctuations at the Gaussian level \cite{de1993crossover,Diener2008}, and therefore is termed as the Gaussian pair fluctuation (GPF) theory. Although in the normal state the modification only leads to fairly small changes \cite{Huicomparitive08}, it yields a non-trivial improvement for the superfluid equation of state \cite{Hu2006,Diener2008} due to the thermodynamic consistency in determining the number density (i.e., the number density calculated by using single-particle Green function or thermodynamic potential should be the same). On the other hand, a fully self-consistent $T$-matrix theory was developed by Haussmann and co-workers \cite{Haussmann2007}, taking advantage of satisfying the Luttinger-Ward (LW) identity. Most recently, a partially self-consistent $T$-matrix theory was also considered by Tajima and co-workers \cite{Tajima2017}. To ensure the crucial thermodynamic consistency in the superfluid phase, they proposed to first determine the chemical potential using the $T$-matrix diagrams, from which all the equations of state are then calculated through various exact thermodynamic relations \cite{Tajima2017}.

In general, there is no \emph{a priori} reason to justify the application of various many-body $T$-matrix theories, although the fully self-consistent LW theory seems to be more preferable in the literature. Naively, to quantitatively evaluate the many-body $T$-matrix theories in the superfluid phase, we may consider two useful criteria: (i) in the unitary limit the theory should provide a better explanation of the benchmark universal equation of state, which becomes available due to the increasingly accurate measurements of the unitary Fermi gas \cite{Nascimbene2010,Horikoshi2010,Ku2012}; and (ii) in the BEC limit the theory should be able to recover a Bogoliubov theory of composite molecules with the correct molecular scattering length $a_{\rm M} \simeq 0.60a_s$ \cite{Petrov2004}. For the first criterion, the self-consistent LW theory is clearly favored. For instance, it predicts a Bertsch parameter $\xi=0.360$ of a zero temperature unitary Fermi gas, which is close to the experimental result in the milestone measurement $\xi \simeq 0.376(4)$ \cite{Ku2012}, contrasted with other many-body $T$-matrix predictions (i.e., $\xi = 0.455$ from the non-self-consistent $T$-matrix \cite{Pieri2004}, $\xi=0.401$ from the GPF \cite{Hu2006,Diener2008}, and $\xi = 0.381$ from the partially self-consistent $T$-matrix \cite{Tajima2017}). However, in the BEC limit the self-consistent LW theory gives an incorrect molecular scattering length $2a_s$ and thereby does not fulfill the second criterion. Actually, almost all the many-body $T$-matrix theories predict the same molecular scattering length $2a_s$ and violate the second criterion. The only exception is the GPF theory, which predicts $a_{\rm M} \simeq 0.57a_s$ \cite{Hu2006,Diener2008}, a value very close to the exact result. The same is also true for a two-dimensional interacting Fermi gas \cite{He2015}. The much-improved molecular scattering length from the GPF theory in both three and two dimensions might be qualitatively understood from the renormalization of the dispersion relation of the gapless Goldstone mode upon the change of energy scale \cite{Salasnich2015}. 

It is desirable to develop better field theoretical approaches, benchmarked against the above-mentioned two criteria. In this respect, the GPF theory could provide an ideal starting point. In a previous work \cite{Mulkerin2016}, going beyond the Gaussian fluctuation approximation (or more broadly the many-body $T$-matrix ladder approximation) has been attempted for a strongly interacting Fermi gas in the \emph{normal} state, motivated by the dimensional $\epsilon$-expansion studied in Refs.~\cite{Nishida2006,Nishida_2007,Son2007,Arnold2007}. One selects the important diagrams, which contribute the most near the upper critical dimension of four. In other words, a weighting factor is \emph{artificially} assigned to the diagrams, according to the powers of $\epsilon = 4-d$, although in the calculation the dimension $d \sim 4$ (i.e., $\epsilon \ll 1$ in the initial assumption) eventually one takes the realistic value of three, leading to $\epsilon=1$. The beyond-GPF theory in Ref. \cite{Mulkerin2016} explored the possibility of including all the diagrams up to the order $\mathcal{O}(\epsilon^2)$ near four dimensions. When applied to calculate the universal equation of state of a unitary Fermi gas above the superfluid transition and compared with the benchmark experimental data \cite{Ku2012}, the beyond-GPF theory shows excellent agreement down to the temperature $0.5T_{\rm F}$, better than the self-consistent LW theory \cite{Mulkerin2016}. At high temperature above Fermi degeneracy, it also provides better agreement with the asymptotically exact third-order quantum virial expansion \cite{Liu2009}.

In this work, we aim to apply the beyond-GPF theory to the broken-symmetry \emph{superfluid} phase and calculate the low-temperature dependence of the chemical potential, energy, and Tan contact parameter throughout the BEC-BCS crossover. We critically examine its applicability at low temperatures by using the latest benchmark experimental measurements and the latest QMC simulations. At essentially zero temperature, we focus on the comparison with the experimental data obtained by Horikoshi \textit{et al.} at the University of Tokyo \cite{Horikoshi2017}. While for the universal equation of state in the unitary limit, we consider the most recent measurement by Li \textit{et al.} at University of Science and Technology of China (USTC), Shanghai Branch \cite{USTC2021}.  In both cases, we find excellent quantitative agreement. Further improvement of our beyond-GPF theory is possible, as the systematic inclusion of large-loop Feynman diagrams at higher orders $\mathcal{O}(\epsilon^n)$ ($n\ge 3$) can be achieved by using, for example, the BDMC technique \cite{VanHoucke2012}. 

The paper is set out as follows. In Sec.~\ref{sec:method} we briefly summarize the theoretical framework used and derive the thermodynamic potential within the beyond-GPF theory. In Sec.~\ref{sec:resultsBECBCS} we describe the thermodynamic properties of the BEC-BCS crossover at very low temperature. In Sec.~\ref{sec:resultsUni} we show  the temperature dependence of the thermodynamic properties at unitarity. Finally, in Sec.~\ref{sec:conc} we summarize our results.

\section{Beyond Gaussian pair fluctuation theory} \label{sec:method}

The theoretical framework of the beyond-GPF theory for an interacting Fermi gas in the normal state has been outlined in the work \cite{Mulkerin2016}. In this section, we briefly review the key ideas of the beyond-GPF theory and extend it to the broken-symmetry superfluid phase. 

A three-dimensional interacting Fermi gas at the BEC-BCS crossover is well described by the single-channel model Hamiltonian density for field operators $\psi_{\sigma}$ \cite{Hu2010},
\begin{alignat}{1} \label{eq:hamil}
\mathcal{H}=\sum_{\sigma}{\psi}^{\dagger}_{\sigma}(x)\mathcal{H}_0\psi_{\sigma}(x)+U_0{\psi}^{\dagger}_{\uparrow}(x){\psi}^{\dagger}_{\downarrow}(x)\psi_{\downarrow}(x)\psi_{\uparrow}(x),
\end{alignat}
where $\mathcal{H}_0=-\nabla^{2}/(2M)-\mu$, $M$ is the mass of fermions, and $\mu$ is the chemical potential.
Throughout the work we shall use the notation $x=(\tau,\mathbf{x})$ and set the reduced Planck constant $\hbar= 1$ and the volume $V=1$. We consider
a contact interaction with strength $U_0<0$, which is known to have an ultraviolet divergence \cite{Bloch2008}. This can be resolved by relating the bare contact interaction strength $U_0$ to the $s$-wave scattering length $a_s$ via,
\begin{equation}
\frac{M}{4\pi a_s} =\frac{1}{U_0}+ \sum_{\mathbf{k}} \frac{1}{\mathbf{k}^2/M}.
\end{equation}

A convenient method to calculate the thermodynamics of ultracold interacting Fermi gases is through the thermodynamic potential: $\Omega=-k_{{\rm B}}T\ln\mathcal{Z}$,
where $T$ is the temperature, $\mathcal{Z}=\int\mathcal{D}\left[\psi,\bar{\psi}\right]e^{-S\left[\psi,\bar{\psi}\right]}$ is the partition function,
and $\psi$ and $\bar{\psi}$ are independent Grassmann fields.
The partition function is defined by the action and Hamiltonian in Eq.~\eqref{eq:hamil} via
\begin{alignat}{1} \label{eq:S}
S[\psi,\bar{\psi}]=\int_{0}^{\beta}d\tau\int d\mathbf{r}\left[\sum_{\sigma}\bar{\psi}_{\sigma}(x)\partial_{\tau}\psi_{\sigma}(x)+\mathcal{H}\right],
\end{alignat}
where $\beta=1/(k_{\rm B}T)$. In the strongly interacting regime we must go beyond the qualitatively reliable mean-field (MF) theory and include higher-order quantum fluctuations.
Taking the Hubbard-Stratonovich transformation 
we write the action
in terms of a bosonic field, $\Delta(\mathcal{Q})$, and
expand about its saddle point $\Delta_{0}$, $\Delta(\mathcal{Q})=\Delta_{0}+\varphi_{\mathcal{Q}}$, where $\varphi_{\mathcal{Q}}$ is the field operator representing fluctuations. Here, we have used for short-hand notation $\mathcal{Q}=(i\nu_n,\mathbf{q})$ and $\sum_{\mathcal{Q}} \equiv (k_BT/V)\sum_{i\nu_{n}}\sum_{\mathbf{q}}$, and $\nu_n=2 n\pi/\beta$ ($n \in \mathbb{Z}$) are the bosonic Matsubara frequencies.
After integrating out the Grassmann fields and taking a perturbative expansion of the bosonic action in orders of
the fluctuation field, we have $\scS_{{\rm eff}}\left[\Delta,\Delta^{*}\right]=\scS_{{\rm MF}}^{(0)}+\scS_{{\rm GF}}^{(2)}+\scS^{(3)}+\scS^{(4)}+\dots$,
where $\scS_{{\rm MF}}^{(0)}$ is the mean-field contribution, $\scS_{{\rm GF}}^{(2)}$ is the Gaussian contribution, and $\scS^{(n)}$ are higher order contributions \cite{Mulkerin2016}. The conceptually simple picture of including fluctuations up to the Gaussian order captures the most relevant physics for strongly-coupled systems, by including the contributions from fermions and Cooper pairs. This GPF model provides a reasonable description of the BEC-BCS crossover in both three dimensions \cite{Hu2006,Hu2007,Diener2008,tempere2012path} and two-dimensions \cite{He2015,Mulkerin2017}. However, in order to find a more accurate description of a strongly interacting Fermi gas, higher-order pair fluctuations beyond the Gaussian level need to be taken into the thermodynamic potential. 

In previous work \cite{Mulkerin2016}, motivated by the dimensional $\epsilon$-expansion \cite{Nishida2006,Arnold2007,Son2007,Nishida_2007} and re-interpreting the small dimensional parameter $\epsilon$ as the pair Green's function, $\Gamma(\mathcal{Q})$, the thermodynamic potential beyond GPF up to order $\mathcal{O}(\epsilon^2)$ was shown to be  $\Omega=\Omega_{\rm MF}+\Omega_{\rm GF} +(\beta V)^{-1}\langle \scS_4\rangle$,  near the upper critical four dimensions. Here, the mean-field thermodynamic potential is given by,
\begin{equation}
\Omega_{\textrm{MF}}=\frac{\Delta^{2}}{U_0}+\sum_{\mathbf{k}}\left[\xi_{\mathbf{k}}-E_{\mathbf{k}}-\frac{2}{\beta}\ln\left(1+e^{\beta E_{\mathbf{k}}}\right)\right],\label{OmegaMF}
\end{equation}
$\xi_{\mathbf{k}}=\mathbf{k}^{2}/(2M)-\mu$,
$E_{\mathbf{k}}=\sqrt{\xi_{\mathbf{k}}^{2}+\Delta^{2}}$,
and to ensure the gapless Goldstone mode, the pairing gap $\Delta$
should be calculated using the mean-field gap equation: 
\begin{alignat}{1}\label{eq:gap}
\frac{m}{4\pi a_s}+\sum_{\mathbf{k}}\left[\frac{1-2f\left(E_{\mathbf{k}}\right)}{2E_{\mathbf{k}}}-\frac{1}{\mathbf{k}^2/M}\right]=0,
\end{alignat}
with the Fermi distribution function $f(x)=\smash{1/(e^{\beta x}+1)}$. A significant advantage of our methodology is that the gap equation is calculated at the mean-field level and satisfies Goldstone's theorem while still including quantum fluctuations. This is in sharp contrast to the self-consistent LW theory, where a running coupling constant should be assumed in order to satisfy the gapless condition \cite{Haussmann2007}. The expression for the GPF thermodynamic potential is determined by integrating out the pairing fluctuation fields from the action, arriving at $\Omega_{\rm GF} = k_{B}T\sum_{\mathcal{Q}}\mathcal{F}\left(\mathcal{Q}\right)e^{i\nu_{n}0^{+}}$, where the function $\mathcal{F}\left(\mathcal{Q}\right) \equiv (1/2) \ln[-\Gamma^{-1}(\mathcal{Q})]$ can be rewritten as 
\begin{alignat}{1}
\mathcal{F}= \frac{1}{2}\ln\left[1-\frac{M_{12}^{2}\left(\mathcal{Q}\right)}{M_{11}\left(\mathcal{Q}\right)M_{11}\left(-\mathcal{Q}\right)}\right]+\ln M_{11}\left(\mathcal{Q}\right).
\end{alignat}
The matrix elements $M_{11}$ and $M_{12}$ of the inverse pair correlation function $\Gamma^{-1}(\mathcal{Q})$ are given in Appendix \ref{Appendixthermo}. 

By extending the work \cite{Mulkerin2016} to the broken-symmetry phase, as detailed in Appendix \ref{Appendixthermo}, the beyond Gaussian pair fluctuation contribution to the thermodynamic potential can be expressed by the three terms: 
\begin{equation}
\frac{\langle \scS^{(4)}\rangle}{\beta V} = \Omega^{(a)}+\Omega^{(b)}+\Omega^{(c)},
\end{equation} 
where,
\begin{alignat}{1}
\Omega^{(a)} & =\sum_{\mathcal{K}}G_{11}^{(0)}(\mathcal{K})G_{11}^{(0)}(\mathcal{K})\Sigma_{11}^{(0)}(\mathcal{K})\Sigma_{11}^{(0)}(\mathcal{K}), \nonumber \\
\Omega^{(b)} & =\sum_{\mathcal{K}}G_{12}^{(0)}(\mathcal{K})G_{12}^{(0)}(\mathcal{K})\Sigma_{11}^{(0)}(\mathcal{K})\Sigma_{22}^{(0)}(\mathcal{K}), \nonumber \\
\Omega^{(c)} & =\sum_{\mathcal{K}\mathcal{Q}_{1}\mathcal{Q}_{2}}G_{12}^{(0)}(\mathcal{K})G_{11}^{(0)}(\mathcal{K}-\mathcal{Q}_{1})G_{12}^{(0)}(\mathcal{K}-\mathcal{Q}_{1}+\mathcal{Q}_{2}) \nonumber \\ 
& \hspace{35pt} \times G_{22}^{(0)}(\mathcal{K}+\mathcal{Q}_{2})\Gamma_{22}(\mathcal{Q}_{1})\Gamma_{22}(\mathcal{Q}_{2}). \label{eq:S4a}
\end{alignat}
Here the BCS Green's function is given by $G^{(0)}(\mathcal{K}) = [i\omega_{m}-\xi_{\mathbf{k}}\tau_3  +\Delta\tau_1]^{-1}$, and $\tau_i$ are the Pauli matrices.
The self-energy $\Sigma^{(0)}$, arising from the beyond GPF diagrams, takes the form \cite{Mulkerin2016},
\begin{alignat}{1}
\Sigma_{\alpha\alpha'}^{(0)}(\mathcal{K}) &= \sum_{\mathcal{Q}} \Gamma_{\alpha\alpha'}(\mathcal{Q}) G_{\alpha\alpha'}^{(0)}(\mathcal{Q}-\mathcal{K}) , 
\end{alignat}
for $\{\alpha,\alpha'\}=\{1,2\}$. We use for short-hand notation $\mathcal{K}=(i\omega_m,\mathbf{k})$ and $\sum_{\mathcal{K}} \equiv (k_BT/V)\sum_{i\omega_{m}}\sum_{\mathbf{k}}$, and $\omega_m=(2m+1)\pi /\beta$ ($m \in \mathbb{Z}$) are the fermionic Matsubara frequencies. 

For the sake of clarity, we leave the technical details of the beyond-GPF calculations to Appendix \ref{Appendixthermo}. In our calculations, the number density is determined through the thermodynamic potential by $n=-\partial\Omega/\partial\mu$, which fixes the Fermi wavevector $k_{\rm F}=(3\pi^2n)^{1/3}$. We self-consistently converge the number density through adjusting the chemical potential and order parameter using Eq.~\eqref{eq:gap}. The temperature scale and energy scale are defined through $k_{\rm B}T_{\rm F}=\epsilon_{\rm F}=k^2_{\rm F}/2M$. The dimensionless interaction parameter is given by $1/(k_{\rm F}a_s)$. 

%%%%%%%%%%%%%%%%%%%%%
\begin{figure}
\includegraphics[width=0.50\textwidth]{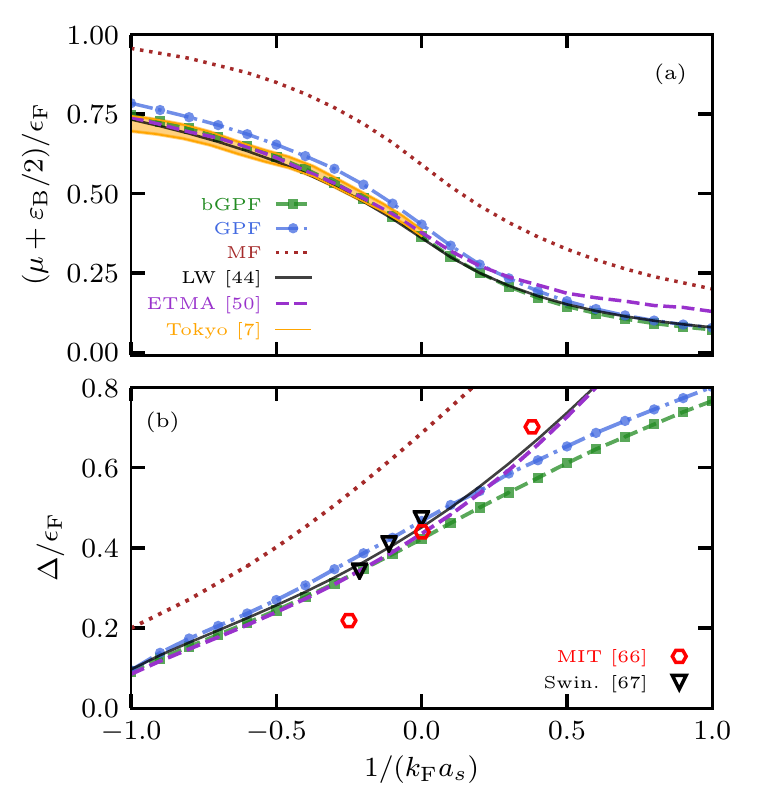} 
\caption{The chemical potential (a) and the superfluid gap parameter (b) in units of the Fermi energy are plotted as a function of the dimensionless interaction parameter $1/(k_{\rm F}a_s)$ at the reduced temperature $T/T_{\rm F}=0.06$. For the chemical potential, we have subtracted half the bound state energy, $-\varepsilon_B/2=-1/(2Ma_s^2)$. We plot the predictions from different theories: mean-field (brown dotted line), GPF (blue line with  circles) \cite{Hu2006}, partially self-consistent or extended $T$-matrix (ETMA, purple dashed line) \cite{Tajima2017}, fully self-consistent LW theory (black solid line) \cite{Haussmann2007}, and beyond-GPF calculation (green line with squares).  We compare the different theoretical predictions with the experimental data from University of Tokyo \cite{Horikoshi2017} (orange solid line; $T/T_{\rm F}\simeq0.06$). For the superfluid gap parameter, we instead compare the theories with the experimental data from MIT \cite{Schirotzek2008} (red hexagons) and Swinburne University of Technology \cite{Hoinka2017} (black triangles; $T/T_{\rm F}\simeq0.09$).}
\label{fig:mu_delta} 
\end{figure}
%%%%%%%%%%%%%%%%%%%%%

\section{BEC-BCS crossover at very low temperatures}\label{sec:resultsBECBCS}
To begin our discussion of the beyond-GPF results, we first consider the chemical potential and superfluid gap parameter at the BEC-BCS crossover as a function of the dimensionless interaction parameter $-1 \leq 1/k_{\rm F}a_s \leq +1$ at the temperature $T/T_{\rm F}=0.06$. This is the temperature set by the experiment on the ground-state thermodynamics at the University of Tokyo \cite{Horikoshi2017}. As in the experiment, we expect the temperature dependence below $T/T_{\rm F}=0.06$ to not be significant and we discuss essentially the ground-state properties of the BEC-BCS crossover. Here, we focus on the comparison of different strong-coupling theories with respect to the benchmark experimental data from Tokyo group \cite{Horikoshi2017}, as reported in Fig.  \ref{fig:mu_delta}.

As we already mentioned in the Introduction, the GPF, extended $T$-matrix ETMA, and fully self-consistent LW theories all belong to the many-body $T$-matrix theories within the ladder approximation, but differ in the degree of self-consistency adopted in the ladder diagrams.  For the chemical potential in Fig.  \ref{fig:mu_delta}(a), in comparison to the benchmark Tokyo data on the BCS side (i.e., $-1 \leq 1/k_{\rm F}a_s \leq 0$), all the $T$-matrix theories show good agreements, much better than the mean-field prediction shown in the brown dotted line. In particular, both the ETMA and LW theories agree  well with the experimental results, with discrepancies smaller than the error bar of data. The simplest non-self-consistent $T$-matrix theory, the GPF theory, seems to provide a worse agreement on the BCS side, as its prediction lies systematically above the experimental data. This weakness can be overcome by including the higher-order diagrams beyond the GPF approximation. Indeed, within our scheme of the inclusion of pair fluctuations at the second order $\mathcal{O}(\epsilon^2)$ in the $\epsilon$-expansion, we find that the beyond-GPF theory agrees with the Tokyo data within the experimental uncertainty. 

On the BEC side of the Feshbach resonance (i.e., $1/k_{\rm F}a_s \geq 0$), recent experimental data is not available. However, it is known analytically that both the ETMA \cite{Tajima2017} and LW theories \cite{Haussmann2007} fail to predict the correct molecular scattering length $a_{\rm M}$ between two composite bosons. This failure can be partly seen from the chemical potential given by the ETMA theory, which slightly shifts up at large $1/(k_Fa_s)$ compared with other theories. In contrast, the GPF theory provides an approximately accurate molecular scattering length $a_{\rm M} \simeq 0.57a_s$ \cite{Hu2006,Diener2008}.  Although we can not obtain analytically the molecular scattering length in the beyond-GPF theory, it is readily seen that, towards the BEC limit both the GPF and beyond-GPF theories nearly predict the same chemical potential. This strongly indicates that the pair fluctuations at the second order $\mathcal{O}(\epsilon^2)$ becomes less important in the BEC limit and therefore does not contribute to the molecular scattering length. As a result, the beyond-GPF theory inherits the accurate molecular scattering length of the GPF theory in the BEC limit. Therefore, among the different strong-coupling theories considered in Fig. \ref{fig:mu_delta}, it seems that the beyond-GPF theory somehow provides \emph{overall} the best prediction at the BEC-BCS crossover for the low-temperature thermodynamics. Further measurements of thermodynamics on the BEC side will be useful to examine this anticipation.

Another useful experimental comparison comes from dynamical measurements of the superfluid gap parameter through the radio-frequency spectroscopy \cite{Schirotzek2008} or Bragg spectroscopy \cite{Hoinka2017}. Fig. \ref{fig:mu_delta}(b) plots the gap parameters predicted by different strong-coupling theories and compares them to the experimental results from MIT \cite{Schirotzek2008} and Swinburne University of Technology \cite{Hoinka2017}. The slight but systematic improvement of the beyond-GPF theory over the simplest GPF theory is clearly visible. On the BCS side, the beyond-GPF prediction agrees well with the results from the ETMA and LW theories. On the BEC side, the discrepancies among the different theories simply follows the different chemical potential as seen from Fig. \ref{fig:mu_delta}(a), since the gap parameter is calculated by using the mean-field gap equation at given chemical potential. The only exception is the fully self-consistent LW theory \cite{Haussmann2007}, where a running coupling constant is used to ensure the gapless condition for the Goldstone phonon mode, so the gap parameter no longer follows the mean-field gap equation. This explains why the ETMA and LW theories predict similar gap parameter on the BEC side, despite the different chemical potential. We note that, at current stage the spectroscopic measurement of the gap parameter in general has less accuracy than the thermodynamic measurement as shown in Fig. \ref{fig:mu_delta}(a). Thus, the comparison of the theories to the experimental data for the gap parameter given in Fig. \ref{fig:mu_delta}(b) is just indicative. We anticipate that future measurement of the gap parameter with novel high-resolution Bragg spectroscopy \cite{USTC2021} may provide us a stringent test of many-body strong-coupling theories.

%%%%%%%%%%%%%%%%%%%%%
\begin{figure}
\includegraphics[width=0.50\textwidth]{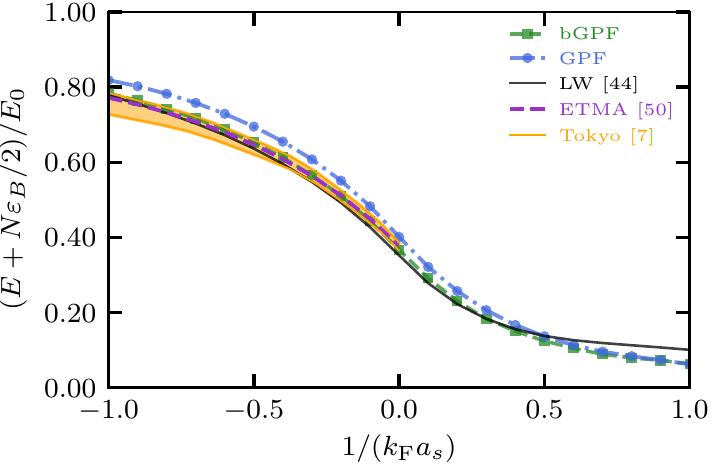} 
\caption{The internal energy in units of $E_0=(3/5)N\epsilon_{\rm F}$ relative to the bound state energy is plotted as a function of the interaction parameter $1/(k_{\rm F}a_s)$ at the reduced temperature $T/T_{\rm F}=0.06$. We plot the energy calculated from the GPF (blue dash-dotted), extended $T$-matrix ETMA (purple dashed), fully self-consistent LW (black solid) and beyond-GPF theories (green dashed). The experimental data on energy is from University of Tokyo in Ref.~\cite{Horikoshi2017} (orange solid line) at the same reduced temperature $T/T_{\rm F}\simeq0.06$.}
\label{fig:En_v}
\end{figure}
%%%%%%%%%%%%%%%%%%%%%

We consider next the internal energy at the BEC-BCS crossover. An advantage of our beyond-GPF calculation is that, once the chemical potential is found for a given interaction strength, we can calculate straightforwardly various thermodynamic quantities. For example, the internal energy is found from the relation:
\begin{alignat}{1}
E = \Omega+TS+\mu N,
\end{alignat}
where $S=(-\partial \Omega/\partial T)_{\mu}$ is the entropy. We plot the internal energy in Fig.~\ref{fig:En_v} in units of  $E_0=(3/5)N\epsilon_{\rm F}$ relative to the bound-state energy of $N/2$ molecules, $-N\varepsilon_B/2=-N/(2Ma_s^2)$,  as a function of $1/k_{\rm F}a_s$ at the temperature $T/T_{\rm F}=0.06$. The predictions from the different theories, the GPF (blue dash-dotted), extended $T$-matrix (purple dashed), fully self-consistent LW (black solid) and beyond-GPF (green solid),  are compared with the experimental data from Ref.~\cite{Horikoshi2017}. Once again, on the BCS side we see the improvement of the GPF prediction with the inclusion of the higher-order non-Gaussian pair fluctuations and the excellent agreement between experiment and theory. Towards the deep BEC limit, the non-Gaussian pair fluctuations become less important and we find that both the GPF and beyond-GPF theories tend to give the same internal energy. The fully self-consistent LW theory predicts a slightly higher internal energy than the beyond-GPF theory in the BEC limit. This is in line with the fact that the LW theory does not give the correct molecular scattering length: it predicts a mean-field molecular scattering length $a_{\rm M}=2a_s$ \cite{Haussmann2007}, about three times larger than the exact value of $a_{\rm M} \simeq 0.6a_s$ \cite{Petrov2004}.

%%%%%%%%%%%%%%%%%%%%%
\begin{figure}
\includegraphics[width=0.50\textwidth]{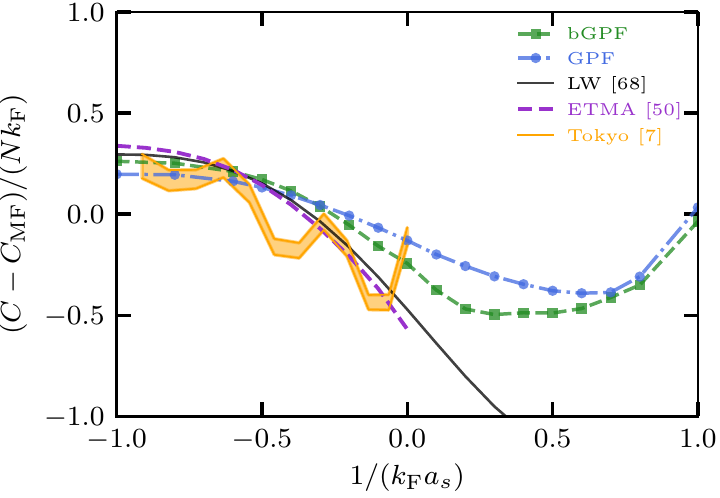}
\caption{Tan's contact parameter in units of $Nk_{\rm F}$ relative to the mean-field prediction is plotted as a function of the interaction parameter $1/(k_{\rm F}a_s)$ at the reduced temperature $T/T_{\rm F}=0.06$. We compare the theoretical predictions from the GPF (blue dash-dotted), extended $T$-matrix ETMA (purple dashed), fully self-consistent LW \cite{haussmann2009} (black solid) and beyond-GPF theories (green dashed), with the experimental results from University of Tokyo in Ref.~\cite{Horikoshi2017} (orange solid line) at the same reduced temperature $T/T_{\rm F}\simeq0.06$.}
\label{fig:En_cont}
\end{figure}
%%%%%%%%%%%%%%%%%%%%%

Another important thermodynamic quantity is Tan's contact parameter $C$, which determines the universal short-range, large-momentum and high-energy behavior of the many-body system \cite{Tan1,Tan2,Tan3}. From the adiabatic relation in the grand canonical ensemble \cite{Hu2011b}, the contact is related to the change in the thermodynamic potential as the interatomic interaction varies:
\begin{alignat}{1} \label{eq:cont}
-\left[\frac{\partial\Omega}{\partial(a_s^{-1})}\right]_{\mu,T} = \frac{C}{4\pi M}.
\end{alignat} 
Figure \ref{fig:En_cont} shows the contact in units of $Nk_{\rm F}$ as a function of the interaction parameter $1/k_{\rm F}a_s$. The mean-field contribution has been subtracted, in order to highlight the difference between theories. For comparison, we show also the experimental data from \cite{Horikoshi2017}. However, this comparison does not serve as an independent benchmark examination of the different theories, since the experimental data are derived from the measured internal energy and are not from other independent spectroscopic measurements such as those in Ref. \cite{Sagi2014}. On the BCS side, there are good agreements between theories, as we already see in the chemical potential, gap parameter and internal energy. On the BEC side, we find that the GPF and beyond-GPF theories consistently predict a larger contact parameter than the LW theory. This is merely a reflection of the difference in the internal energy that we have discussed earlier. 
 
%%%%%%%%%%%%%%%%%%%%%
\begin{figure}
\includegraphics[width=0.52\textwidth]{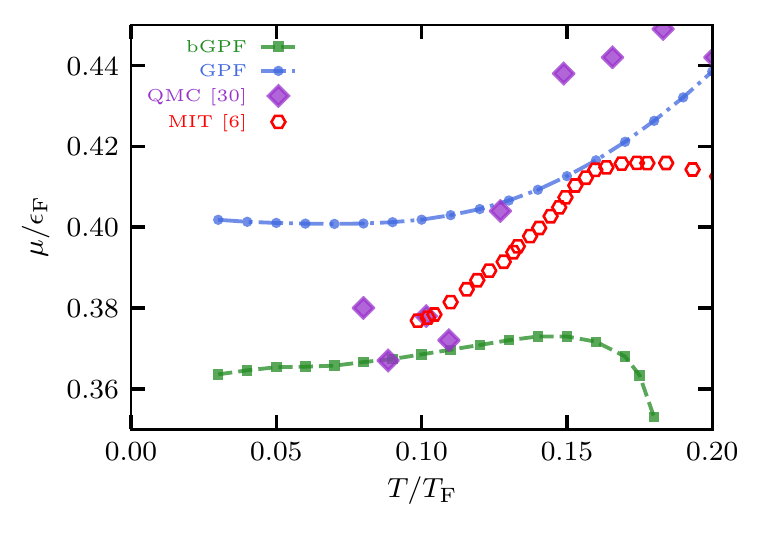} 
\caption{The chemical potential in units of the Fermi energy $\epsilon_{\rm F}$ is plotted as a function of the reduced temperature in the unitary limit, $1/(k_{\rm F}a_s)=0$. These predictions are calculated by using the GPF theory (blue dash-dotted), beyond-GPF theory (green dashed), and lattice QMC from Ref.~\cite{Goulko2016QMC} (purple diamonds). We compare the theoretical results with the experimental data from MIT in Ref.~\cite{Ku2012} (red hexagons).}
\label{fig:unitary_mu} 
\end{figure}
%%%%%%%%%%%%%%%%%%%%%

\section{Universal thermodynamics of a unitary Fermi gas} \label{sec:resultsUni}
We now turn our attention to the temperature dependence of thermodynamic properties of a unitary Fermi gas and fix the $s$-wave scattering length to the unitary limit $1/(k_{\rm F}a_s)=0$. In this limit, the interacting Fermi gas becomes scale invariant and all the thermodynamic functions can be expressed as a single function of the reduced temperature $T/T_F$. This unique advantage greatly simplifies the measurements of the equations of state and also improves the accuracy of experimental data. Universal thermodynamic function of the unitary Fermi gas have now been measured with unprecedented accuracy \cite{Ku2012,USTC2021}, with the characteristic lambda transition clearly revealed in the specific heat. Here, we focus on the comparison of the GPF and beyond-GPF theories with the latest experiment carried out at USTC Shanghai \cite{USTC2021}. This measurement benefits from a large homogeneous Fermi gas trapped in a box-potential and a novel high-resolution Bragg spectroscopy for the isentropic speed of sound \cite{USTC2021}. As a result, the experiment removes the local density approximation assumed in the previous measurement \cite{Ku2012}. We also consider the comparison of the GPF and beyond-GPF theories with the latest available QMC simulations \cite{Goulko2016QMC,Jensen2020}.

We first plot the chemical potential as a function of the reduced temperature at unitarity in Fig.~\ref{fig:unitary_mu}, which is calculated using the GPF (blue dash-dotted) and beyond-GPF theories (green dashed). As we approach zero temperature the Fermi surface becomes sharp, and in our numerical calculations accuracy becomes worse at very low temperatures. Therefore, we restrict ourselves to temperatures $T/T_{\rm F}\geq 0.03$. We compare the two theoretical predictions with the lattice QMC in Ref.~\cite{Goulko2016QMC} (purple diamonds) and the experimental chemical potential measured in Ref.~\cite{Ku2012}.

At low temperature $T/T_{\rm F}<0.1$, the GPF theory overestimates the chemical potential. We find instead that the beyond-GPF prediction is consistent with either the experimental result of Ref.~\cite{Ku2012} or the finite-temperature QMC simulation \cite{Goulko2016QMC}. However, if we further increase temperature, the beyond-GPF theory seems to break down, presumably due to large critical thermal fluctuations close to the superfluid phase transition. This is clearly indicated by a maximum at $T/T_{\rm F} \simeq 0.15$ in the beyond-GPF chemical potential. The chemical potential continues to decrease for higher temperatures. We are not able to determine the superfluid transition temperature within the beyond-GPF theory. The beyond-GPF chemical potential shown in Fig.~\ref{fig:unitary_mu} is too large to yield a vanishing gap parameter. To locate the superfluid transition temperature, it might be useful to calculate the superfluid density using the beyond-GPF theory in future work.  

%%%%%%%%%%%%%%%%%%%%%
\begin{figure}
\includegraphics[width=0.50\textwidth]{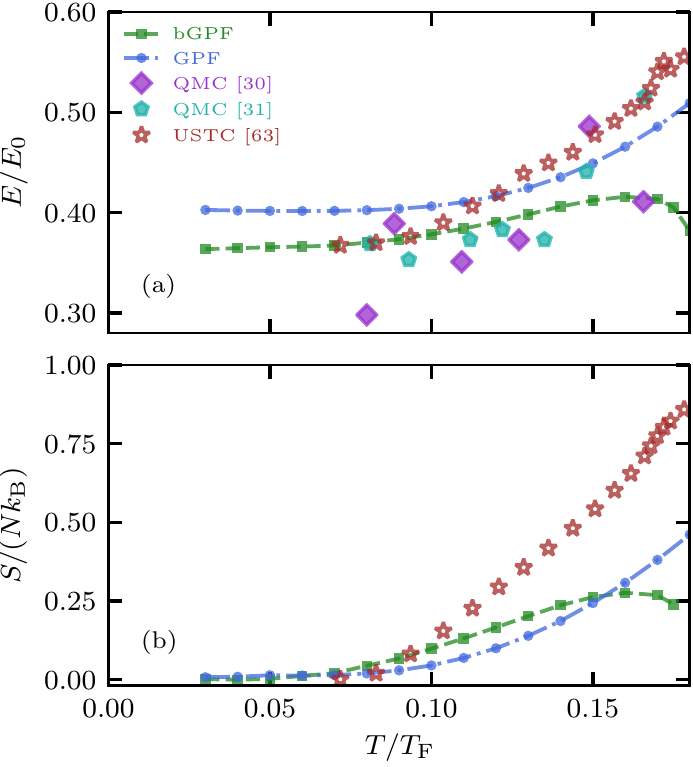} 
\caption{(a) The internal energy of a unitary Fermi gas is plotted as a function of the reduced temperature in units of an ideal gas energy, $E_0=(3/5)N\epsilon_{\rm F}$. We show the predictions from the GPF (blue circles), beyond-GPF (green squares), lattice QMC from Ref.~\cite{Goulko2016QMC} (purple diamonds) and auxiliary-field QMC from Ref.~\cite{Jensen2020} (light-blue pentagons). We also compare the predictions with the experimental results from UTSC-Shanghai (brown stars) \cite{USTC2021}. (b) The entropy of a unitary Fermi gas is plotted as a function of reduced temperature in units $Nk_{\rm B}$ for the GPF (blue circles) and beyond-GPF calculations (green squares), and the experimental data from UTSC-Shanghai (brown stars) \cite{USTC2021}.}
\label{fig:unitary_S_E} 
\end{figure}
%%%%%%%%%%%%%%%%%%%%%

We now consider the internal energy, which at unitarity obeys the well-known scaling relation $E=-2\Omega/3$  \cite{Ho2004,Nascimbene2010}. In Fig.~\ref{fig:unitary_S_E}(a) we plot the internal energy in units of the ideal-gas energy at zero temperature $E_0=(3/5)N\epsilon_{\rm F}$ as a function of the reduced temperature. We compare the beyond-GPF energy (green dashed) to the theoretical predictions of the GPF theory (blue dash-dotted), lattice QMC results from Ref.~\cite{Goulko2016QMC} (purple diamonds) and auxiliary-field QMC results from Ref.~\cite{Jensen2020} (light-blue pentagons), and experimental data from UTSC Shanghai \cite{USTC2021} (brown stars). The internal energy calculated from our beyond-GPF theory agrees well with both QMC results and the experimental data in the low temperature regime. In particular, it is remarkable that at temperatures $T/T_{\rm F} < 0.1$ the beyond-GPF energy exactly overlaps with the experimental data. The unitary Fermi gas can be related to the ideal Fermi gas at zero temperature via the Bertsch parameter, i.e., $E(T=0)=\xi E_0$. We find that the beyond-GPF theory predicts a Bertsch parameter $\xi \simeq 0.365$, in excellent agreement with the zero-temperature QMC $\xi = 0.367(7)$ found in Ref.~\cite{Jensen2020} and the experimental measurement $\xi = 0.367(9)$ in Ref.~\cite{USTC2021}. As the temperature increases above $T/T_{\rm F} = 0.1$, we see the beyond-GPF energy starts to deviate from the benchmark experimental data. Indeed, near the critical temperature $T_c \simeq 0.17T_{\rm F}$, accurate theoretical predictions of the internal energy are notoriously difficult to obtain. This is partly indicated by the uncertainty of the two QMC results. Therefore, in order to accurately account for the experimental data, even the most advanced QMC simulations need to be improved.

In Fig. \ref{fig:unitary_S_E}(b) we report the entropy in units of $Nk_{\rm B}$ as a function of the reduced temperature at unitarity. In the low temperature regime (i.e., $T/T_{\rm F}<0.1$), once again we find the excellent agreement between the beyond-GPF theory and the experiment. As the temperature increases, however, the beyond-GPF theory turns out to strongly under-estimate the entropy.

%%%%%%%%%%%%%%%%%%%%%
\begin{figure}
\includegraphics[width=0.50\textwidth]{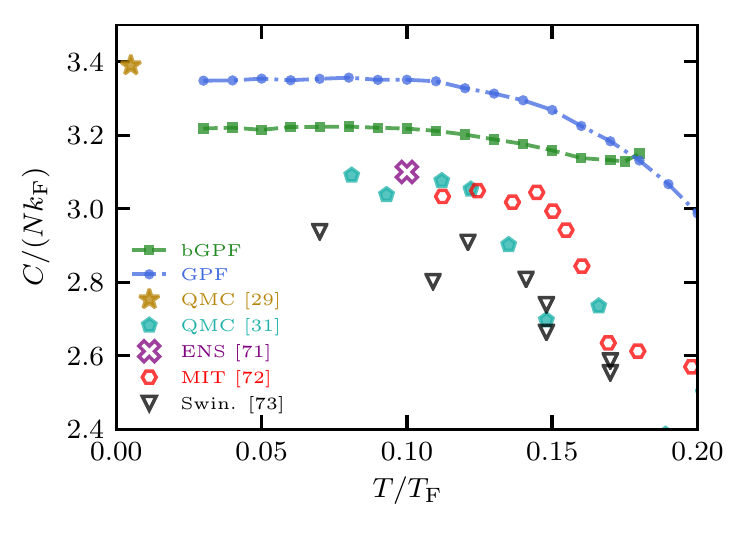} 
\caption{ Tan's contact parameter of a unitary Fermi gas in units of $Nk_{\rm F}$ is plotted as a function of the reduced temperature for the GPF (blue dash-dotted) and beyond-GPF calculations (green dashed), finite-temperature auxiliary-field QMC \cite{Jensen2020} (light-blue pentagons) and zero-temperature diffusion QMC simulations \cite{Carlson2011} (golden star). For comparison, we show also the early experimental result from Ref.~\cite{Laurent2017} (purple cross) and the latest measurements from Ref.~\cite{Mukherjee2019} (red hexagons) and Ref.~\cite{Carcy2019} (black triangles).}
\label{fig:unitary_cont} 
\end{figure}
%%%%%%%%%%%%%%%%%%%%%

Finally, we show in Fig.~\ref{fig:unitary_cont} Tan's contact parameter in units of $Nk_{\rm F}$ as a function of the reduced temperature and compare the predictions of the GPF and beyond-GPF theories to several QMC calculations and experimental results. Our new calculation of the beyond-GPF theory (green dashed) systematically improves the earlier GPF result in Ref. \cite{Hu2011b}. As a consequence, in the low temperature regime the beyond-GPF prediction now seems to be able to smoothly connect to the various experimental data, from \'{E}cole Normale Sup\'{e}rieure (ENS) \cite{Laurent2017} (brown cross), MIT  \cite{Mukherjee2019} (red hexagons) and Swinburne University of Technology \cite{Carcy2019} (black triangle). As the critical temperature is approached, the contact parameter given by the beyond-GPF theory does not decrease as sharply as the experimental results, since the theory fails to include strong thermal fluctuations in this critical temperature regime.

It is evident from Figs. \ref{fig:unitary_mu}, \ref{fig:unitary_S_E} and \ref{fig:unitary_cont} that, although the beyond-GPF theory does not agree with the experimental data close to the superfluid transition, it is able to provide a \emph{quantitative} explanation of the experimental results below the temperature $T<0.1T_{\rm F} \simeq 0.6T_c$. This accurate low-temperature description might be useful in understanding the thermodynamics of strongly interacting fermions at the intermediate low temperature regime, which may become difficult to address by both experiments (due to the less efficient thermometry) and QMC simulations (due to the rapid increase in simulation time at low temperatures). 

\section{Conclusion} \label{sec:conc}
In summary, we have performed numerical calculations of the thermodynamics of a strongly interacting Fermi superfluid in the broken-symmetry phase, by using a recently developed strong-coupling theory that includes pair fluctuations beyond the standard Gauss approximation \cite{Mulkerin2016}. We have critically examined the applicability of the beyond-GPF (Gaussian pair fluctuation) theory, by comparing its predictions for various equations of state and Tan's contact parameter with the most recent experimental measurements and the latest quantum Monte Carlo simulations. On the BCS side and in the unitary limit, we find that the beyond-GPF theory is quantitatively reliable at low temperature. In particular, in the unitary limit, it predicts very a accurate chemical potential, energy, entropy and contact parameter. The beyond-GPF also leads to a Bertsch parameter that is in excellent agreement with the latest experiment \cite{USTC2021} and Monte Carlo calculation \cite{Jensen2020}. It is reasonable to believe that the beyond-GPF theory is also quantitatively reliable on the BEC side at low temperature and future experimental measurements on the low-temperature thermodynamics in the BEC regime would be useful to check this anticipation. 

By combining the previous results for a normal, strongly interacting Fermi gas \cite{Mulkerin2016}, we conclude that the beyond-GPF theory is a powerful tool in either low-temperature regime or high-temperature regime. In the former case, quantum fluctuations are stabilized by the pairing gap parameter; while in the latter, thermal fluctuations are significantly suppressed. In the unitary limit, the two temperature regimes are roughly given by $T<0.1T_{F}\simeq0.6T_{c}$ and $T>0.5T_{F}$, respectively, where $T_{F}$ is Fermi degenerate temperature and $T_{c}\simeq0.17T_{F}$ is the superfluid transition temperature at unitarity.

In the intermediate temperature regime (i.e., $0.1T_{F}\leq T\leq0.5T_{F}$ for a unitary Fermi gas), the beyond-GPF theory needs to be improved to capture the strong thermal fluctuations near the superfluid transition. As the current beyond-GPF framework is built on the dimensional $\epsilon$-expansion, an immediate extension of our work is to carry out the beyond-GPF calculations near four dimensions with $d=4-\epsilon$, where the small parameter $0<\epsilon\ll1$ and the large thermal fluctuations near the critical temperature can therefore be suppressed. We may then obtain the universal thermodynamic function such as $f_{E}\left(T/T_{F}\right)=E/E_{0}$ up to the second order $\mathcal{O}(\epsilon^{2})$. This naturally extends the earlier zero-temperature second-order $\epsilon$-expansion calculations in Ref. \cite{Arnold2007} to finite temperatures. In more general terms, we may pick the Feynman diagrams at higher orders $\mathcal{O}(\epsilon^{n})$ with $n\geq3$. The systematic inclusion of the higher-order diagrams may greatly shrink the temperature window near the superfluid transition, where the beyond-GPF theory fails.

\begin{acknowledgments}
We are pleased to acknowledge Munekazu Horikoshi, Hiroyuki Tajima, Olga Goulko, Scott Jensen, Yoram Alhassid, Joaqu\'{\i}n Drut, Gabriel Wlaz{\l}owski, Rudolf Haussman, Chris Vale and Martin Zwierlein for sharing their data. Our research was supported by Australian Research Council's (ARC) Discovery Projects under Grant No. DP180102018 (X.-J.L).

\end{acknowledgments}

\appendix

\section{Calculation details on the beyond Gaussian pair fluctuation theory} \label{Appendixthermo}

In this appendix, we present the full matrix elements for the GPF theory and technical details on the GPF and beyond-GPF calculations.

\subsection{GPF}
The thermodynamic potential including up to Gaussian fluctuations is $\Omega=\Omega_{\rm MF}+\Omega_{\rm GF}$ \cite{Hu2006,Diener2008}, where $\Omega_{\rm MF}$ is the mean-field contribution and $\Omega_{\rm GF} = k_{B}T\sum_{\mathcal{Q}}\mathcal{F}\left(\mathcal{Q}\right)e^{i\nu_{n}0^{+}}$ the Gaussian fluctuation contribution, where 
\begin{alignat}{1}
\mathcal{F}\left(\mathcal{Q}\right) = \frac{1}{2}\ln\left[1-\frac{M_{12}^{2}\left(\mathcal{Q}\right)}{M_{11}\left(\mathcal{Q}\right)M_{11}\left(-\mathcal{Q}\right)}\right]+\ln M_{11}\left(\mathcal{Q}\right).
\end{alignat}
Here, $\mathcal{Q}=(i\nu_n,\mathbf{q})$ and $M_{11}(\mathcal{Q})$ and $M_{12}(\mathcal{Q})$ are the matrix elements of the inverse pair Green function $\Gamma^{-1}(\mathcal{Q})$ at finite temperatures and take the form \cite{Hu2006,Diener2008},
\begin{widetext}
\begin{eqnarray} 
M_{11}(\mathcal{Q}) & = & \frac{1}{U_0}+\sum_{\mathbf{k}}\left[u_{+}^{2}u_{-}^{2}\frac{1-f_{+}-f_{-}}{i{\nu}_{n}-E_{+}-E_{-}}-u_{+}^{2}v_{-}^{2}\frac{f_{+}-f_{-}}{i\nu_{n}-E_{+}+E_{-}}+v_{+}^{2}u_{-}^{2}\frac{f_{+}-f_{-}}{i\nu_{n}+E_{+}-E_{-}}-v_{+}^{2}v_{-}^{2}\frac{1-f_{+}-f_{-}}{i{\nu}_{n}+E_{+}+E_{-}}\right],\nonumber \\
M_{12}(\mathcal{Q}) & = & \sum_{\mathbf{k}}\left(u_{+}v_{+}u_{-}v_{-}\right)\left[-\frac{1-f_{+}-f_{-}}{i\nu_{n}-E_{+}-E_{-}}-\frac{f_{+}-f_{-}}{i\nu_{n}-E_{+}+E_{-}}+\frac{f_{+}-f_{-}}{i\nu_{n}+E_{+}-E_{-}}+\frac{1-f_{+}-f_{-}}{i\nu_{n}+E_{+}+E_{-}}\right],\label{M11M12}
\end{eqnarray}
\end{widetext}
where $M_{21}(\mathcal{Q})=M_{12}(\mathcal{Q})$ and $M_{22}(\mathcal{Q})=M_{11}(-\mathcal{Q})$. We note that, the short-hand notations $E_{\pm}= E_{\mathbf{k}\pm\mathbf{q}/2}$, $f_{\pm}^{\left(\pm\right)}=f(E_{\mathbf{k}\pm\mathbf{q}/2}^{\pm})$,
$u_{\pm}^{2}=(1+\xi_{\mathbf{k}\pm\mathbf{q}/2}/E_{\mathbf{k}\pm\mathbf{q}/2})/2$,
and $v_{\pm}^{2}=1-u_{\pm}^{2}$ are used. 

Although it is straight forward to write down the GPF thermodynamic potential, in general $\Omega_{\rm GF}$ is difficult
to solve numerically at finite temperatures. The bosonic Matsubara frequency sum $\sum_{i\nu_n}  \mathcal{F}\left(i\nu_{n},\mathbf{q}\right)$ is divergent and we need to add an infinitesimal controlling factor $e^{i\nu_{n}\eta}$ with $\eta \rightarrow 0^{+}$ to ensure the convergence. At zero temperature, we may rewrite the summation into an integral, due to the fact that the vertex function $\Gamma(\mathcal{Q})$ does not have poles on the real axis for positive values \cite{Diener2008}. However, this is no longer true at finite temperatures. If one uses the analytic continuation as in the seminal work \cite{nozieres1985bose}, the numerical calculation then becomes extremely complicated \cite{Hu2006}. The numerical difficulty can be avoided by splitting the Matsubara sum into two parts \cite{Mulkerin2017,tempere_d_wave},
\begin{alignat}{1} \label{eq:tempere_split}
\sum_{n}  \mathcal{F}_{\eta}\left(i\nu_{n},\mathbf{q}\right) = \sum_{\left|n\right|>n_0}  \mathcal{F}_{\eta}\left(i\nu_{n},\mathbf{q}\right) +\sum_{n=-n_0}^{n_0}  \mathcal{F}_{\eta}\left(i\nu_{n},\mathbf{q},\right)
\end{alignat}
for \emph{arbitrary} positive integer $n_{0}$. Here, $\mathcal{F}_{\eta}(i\nu_{n},\mathbf{q})\equiv\mathcal{F}(i\nu_{n},\mathbf{q})e^{i\nu_{n}\eta}$. We simplify the first divergent term by closing the contour at finite imaginary constant, off the real axis, i.e. for finite $\gamma=(2n_{0}+1)\pi/\beta$,
\begin{equation}
\frac{1}{\beta}\sum_{\left|n\right|>n_{0}}\mathcal{F}_{\eta}\left(i\nu_{n},\mathbf{q}\right)=-\frac{1}{\pi}\int_{-\infty}^{+\infty}d\omega\frac{\textrm{Im}\mathcal{F_{\eta}}\left(\omega+i\gamma,\mathbf{q}\right)}{e^{\beta\omega}+1}.\label{TempereTrick}
\end{equation}
The second term is convergent and can be calculated directly at bosonic Matsubara frequencies $i\nu_n$.
The contribution to $\Omega_{\textrm{GF}}$ at a given wavevector $\mathbf{q}$
can then be obtained by using Eqs.~(\ref{eq:tempere_split}) and \eqref{TempereTrick}. We have checked that this numerical
method is consistent with previous theoretical calculations in Ref. \cite{Hu2006} and is independent of the choice of $n_{0}$, as it should be.

\subsection{Beyond GPF}
The expression of the beyond Gaussian fluctuation contribution to the thermodynamic potential is determined with the method detailed in Ref.~\cite{Mulkerin2016}. Re-interpreting the $\epsilon$-expansion found by Refs. \cite{Nussinov2006,Nishida2006,Nishida_2007} in orders of the vertex function $\Gamma$, the $\scS^{(4)}$ contribution to the thermodynamic potential is 
\begin{widetext}
\begin{alignat}{1}  \label{eq:longS4}
\frac{\scS^{(4)}}{\beta V} = \frac{1}{4} \sum \mathrm{Tr}_{\sigma}  {G}_0(\mathcal{K}) \fourvec{0}{\varphi_{\mathcal{Q}_1}}{\varphi_{-\mathcal{Q}_1}^*}{0}  {G}_0(\mathcal{K}-\mathcal{Q}_1) \fourvec{0}{\varphi_{-\mathcal{Q}_2}}{\varphi_{\mathcal{Q}_2}^*}{0} 
{G}_0(\tilde{\mathcal{K}}) \fourvec{0}{\varphi_{\mathcal{Q}_3}}{\varphi_{-\mathcal{Q}_3}^*}{0} {G}_0(\tilde{\mathcal{K}}-\mathcal{Q}_3) \fourvec{0}{\varphi_{-\mathcal{Q}_4}}{\varphi_{\mathcal{Q}_4}^*}{0},
\end{alignat}
where the summation is over $\mathcal{K},\mathcal{Q}_1,\mathcal{Q}_2,\mathcal{Q}_3,\mathcal{Q}_4$ with $\mathcal{Q}_1+\mathcal{Q}_2=\mathcal{Q}_3+\mathcal{Q}_4$ and $\tilde{\mathcal{K}} = \mathcal{K}-\mathcal{Q}_1+\mathcal{Q}_2$,  
and the two by two BCS Green's functions are
\begin{alignat}{1}
& G_{11}^{(0)}(i\omega_m,\mathbf{k}) = \frac{u_{\mathbf{k}}^2}{i\omega_m - E_{\mathbf{k}}} + \frac{v_{\mathbf{k}}^2}{i\omega_m + E_{\mathbf{k}}} = -G_{22}^{(0)}(-i\omega_m,-\mathbf{k}) \label{eq:green1} \\
& G_{12}^{(0)}(i\omega_m,\mathbf{k}) = \frac{u_{\mathbf{k}}v_{\mathbf{k}}}{i\omega_m - E_{\mathbf{k}}} + \frac{u_{\mathbf{k}}v_{\mathbf{k}}}{i\omega_m + E_{\mathbf{k}}} = G_{21}^{(0)}(i\omega_m,\mathbf{k}). \label{eq:green2}
\end{alignat}
Expanding Eq.~\eqref{eq:longS4} out (which is a lengthy but straightforward calculation), there are 16 terms in total. Using Wick's theorem to expand terms of the form $\langle  \varphi_{-\mathcal{Q}_1}^*\varphi_{\mathcal{Q}_2}^*\varphi_{-\mathcal{Q}_3}^*\varphi_{\mathcal{Q}_4}^*   \rangle$ with the definitions \cite{Mulkerin2016},
\begin{alignat}{1} 
& \langle \varphi^{*}_\mathcal{Q} \varphi_{\mathcal{Q}'}\rangle  =  \Gamma_{11}(\mathcal{Q})\delta_{\mathcal{Q}\mathcal{Q}'},  \,\,\,
  \langle \varphi^{*}_\mathcal{Q} \varphi^{*}_{\mathcal{Q}'}\rangle  =  \Gamma_{12}(\mathcal{Q})\delta_{\mathcal{Q}-\mathcal{Q}'} \label{Eq:Pair_prop}
\end{alignat}
we have terms like
\begin{alignat}{1}
\langle  \varphi_{-\mathcal{Q}_1}^*\varphi_{\mathcal{Q}_2}^*\varphi_{-\mathcal{Q}_3}^*\varphi_{\mathcal{Q}_4}^*   \rangle = \delta_{\mathcal{Q}_1\mathcal{Q}_2}\delta_{\mathcal{Q}_3\mathcal{Q}_4}\Gamma_{12}(\mathcal{Q}_1)\Gamma_{12}(\mathcal{Q}_3) + \delta_{\mathcal{Q}_1-\mathcal{Q}_3}\delta_{\mathcal{Q}_2-\mathcal{Q}_4}\Gamma_{12}(\mathcal{Q}_1)\Gamma_{12}(\mathcal{Q}_2) + \delta_{\mathcal{Q}_1\mathcal{Q}_3}\delta_{\mathcal{Q}_2\mathcal{Q}_4}\Gamma_{12}(\mathcal{Q}_1)\Gamma_{12}(\mathcal{Q}_2).
\end{alignat}
\end{widetext}
Using the fact that within the re-interpreted $\epsilon$ expansion the pair vertex function $\Gamma_{12}\propto\epsilon^2$ is a higher-order term than $\Gamma_{11} \propto \epsilon$, we may neglect the terms that contain $\Gamma_{12}$ and obtain the three leading contributions \cite{Mulkerin2016}, i.e.,
\begin{equation}
\frac{\langle \scS^{(4)}\rangle}{\beta V} = \Omega^{(a)}+\Omega^{(b)}+\Omega^{(c)},
\end{equation} 
where,
\begin{alignat}{1}
\Omega^{(a)} & =\sum_{\mathcal{K}}G_{11}^{(0)}(\mathcal{K})G_{11}^{(0)}(\mathcal{K})\Sigma_{11}^{(0)}(\mathcal{K})\Sigma_{11}^{(0)}(\mathcal{K}), \nonumber \\
\Omega^{(b)} & =\sum_{\mathcal{K}}G_{12}^{(0)}(\mathcal{K})G_{12}^{(0)}(\mathcal{K})\Sigma_{11}^{(0)}(\mathcal{K})\Sigma_{22}^{(0)}(\mathcal{K}), \nonumber \\
\Omega^{(c)} & =\sum_{\mathcal{K}\mathcal{Q}_{1}\mathcal{Q}_{2}}G_{12}^{(0)}(\mathcal{K})G_{11}^{(0)}(\mathcal{K}-\mathcal{Q}_{1})G_{12}^{(0)}(\mathcal{K}-\mathcal{Q}_{1}+\mathcal{Q}_{2}) \nonumber \\ 
& \hspace{35pt} \times G_{22}^{(0)}(\mathcal{K}+\mathcal{Q}_{2})\Gamma_{22}(\mathcal{Q}_{1})\Gamma_{22}(\mathcal{Q}_{2}),
\end{alignat}
as shown in the main text. Physically, keeping only terms involving  $\Gamma_{11}\Gamma_{11}$ is justified as these involve the scattering processes of two pairs, whereas terms containing $\Gamma_{12}\Gamma_{12}$ involve the scattering processes of four pairs and can be neglected.

To obtain the thermodynamic potential, we need to calculate the self-energy ($\{\alpha,\alpha'\}=\{1,2\}$), 
\begin{alignat}{1}
\Sigma_{\alpha\alpha'}^{(0)}(\mathcal{K}) = \sum_{\mathcal{Q}}  \Gamma^{\,}_{\alpha\alpha'}(\mathcal{Q}) G_{\alpha\alpha'}^{(0)}(\mathcal{Q}-\mathcal{K}).
\end{alignat}
For this purpose, we write out explicitly:
\begin{widetext}
\begin{alignat}{1}
\Sigma_{\alpha\alpha'}^{(0)}(i\omega_m,\mathbf{k}) = \sum_{i\nu_n} \int \frac{d^3\mathbf{q}}{(2\pi)^3} \Gamma_{\alpha\alpha'}(i\nu_n,\mathbf{q}) G_{\alpha\alpha'}^{(0)}(i\nu_n-i\omega_m,\mathbf{q}-\mathbf{k}),
\end{alignat}
with
\begin{alignat}{1}
\Gamma(i\nu_n,\mathbf{q}) = \frac{1}{M_{11}M_{22}-M_{12}M_{21}} \fourvec{M_{22}}{M_{12}}{M_{21}}{M_{11}} = 
\fourvec{\Gamma_{11}(i\nu_n,\mathbf{q})}{\Gamma_{12}(i\nu_n,\mathbf{q})}{\Gamma_{21}(i\nu_n,\mathbf{q})}{\Gamma_{22}(i\nu_n,\mathbf{q})}.
\end{alignat}
It is well known that the Matsubara summation is slowly convergent and several methods have been used to find the self-energy (see Refs.~\cite{Pieri2004,Haussmann2007,Tajima2017}).  Here, we circumvent this problem by closing the contour of the bosonic Matsubara summation off the real axis, in the complex plane \cite{tempere_d_wave,Mulkerin2017}, avoiding directly performing the bosonic Matsubara summation or dealing with the complicated behavior on the real axis after analytical continuation. This contour integration yields
\begin{alignat}{1}
\Sigma_{\alpha\alpha'}^{(0)}&(i\omega_m,\mathbf{k}) =  \frac{1}{2\pi i}\int_{-\infty}^{\infty} d\epsilon \int \frac{d^3\mathbf{q}}{(2\pi)^3} \biggl[ b(\epsilon+i\gamma+i\omega_m)\Gamma_{\alpha\alpha'}(\epsilon+i\gamma+i\omega_m,\mathbf{q}) G_{\alpha\alpha'}^{(0)}(\epsilon+i\gamma,\mathbf{q}-\mathbf{k}) + \nonumber \\ 
& b(\epsilon+i\gamma)\Gamma_{\alpha\alpha'}(\epsilon+i\gamma,\mathbf{q}) G_{\alpha\alpha'}^{(0)}(\epsilon+i\gamma-i\omega_m,\mathbf{q}-\mathbf{k}) +(\gamma \rightarrow-\gamma)\biggl] + \frac{1}{\beta}\sum_{n=n_0}^{n_0}\Gamma_{\alpha\alpha'}(i\nu_n,\mathbf{q})G_{\alpha\alpha'}^{(0)}(i\nu_n-i\omega_m,\mathbf{q}-\mathbf{k}),
\end{alignat}
\end{widetext}
where $\gamma$ is finite and satisfies $\nu_{n_0}<\gamma<\nu_{n_0+1}$ and $\gamma\neq \omega_m$ and the Bose distribution is $b(x) = (e^{\beta x}-1)^{-1}$. 
In practice we use a value of $n_0=0$ and $\gamma=3\pi/4$, in units of inverse temperature. We have confirmed that this numerical procedure is robust and converges independent of $n_0$.  

Although there is no explicit Lehmann representation of $\Gamma_{11}$ it can be shown that the spectral representation $\Gamma_{11}(i\nu_n\rightarrow z,\mathbf{q})$ is analytic off the real axis for complex variable $z$ \cite{Pieri2004}. This ensures that the above procedure of contour integration avoids any poles off the real axis. 

Once the self-energy has been tabulated as a function of the wavevector $|\mathbf{k}|$ and fermionic Matsubara frequency $i\omega_m$, the convergent sums in $\Omega^{(a)} $ and $\Omega^{(b)} $ are computed. The contribution $\Omega^{(c)}$ contains  a six-dimensional integral after  integrating out the frequency part, which is computationally difficult. We simplify the expression of $\Omega^{(c)} $ by inserting the following identity,
\begin{alignat}{1}
\sum_{\tilde{\mathcal{K}}}\int dx \exp\left[ i\left( \tilde{\mathcal{K}}+\mathcal{K}+\mathcal{Q}_1-\mathcal{Q}_2 \right) x \right] = 1
\end{alignat}
and introducing the following function
\begin{alignat}{1}
\Sigma_{11}^{(0)}(\mathcal{K},x) = \sum_{\mathcal{Q}}  e^{i\mathcal{Q}x}\Gamma^{\,}_{11}(\mathcal{Q}) G_{11}^{(0)}(\mathcal{Q}-\mathcal{K}),
\end{alignat}
where $x=\left(\tau,\mathbf{x}\right)$.
After some further manipulations we arrive at
\begin{alignat}{1}
\Omega^{(c)} & = -\int dx \sum_\mathcal{K} G_{12}^{(0)}(\mathcal{K}) G_{12}^{(0)}(x)  \Sigma_{11}^{(0)}(-\mathcal{K},x)  \Sigma_{11}^{(0)}(\mathcal{K},-x).
\end{alignat}
This final equation only requires a three-dimensional integration and single sum over the remaining fermionic Matsubara frequency.

\bibliography{Biblio}

\begin{thebibliography}{75}%
\makeatletter
\providecommand \@ifxundefined [1]{%
 \@ifx{#1\undefined}
}%
\providecommand \@ifnum [1]{%
 \ifnum #1\expandafter \@firstoftwo
 \else \expandafter \@secondoftwo
 \fi
}%
\providecommand \@ifx [1]{%
 \ifx #1\expandafter \@firstoftwo
 \else \expandafter \@secondoftwo
 \fi
}%
\providecommand \natexlab [1]{#1}%
\providecommand \enquote  [1]{``#1''}%
\providecommand \bibnamefont  [1]{#1}%
\providecommand \bibfnamefont [1]{#1}%
\providecommand \citenamefont [1]{#1}%
\providecommand \href@noop [0]{\@secondoftwo}%
\providecommand \href [0]{\begingroup \@sanitize@url \@href}%
\providecommand \@href[1]{\@@startlink{#1}\@@href}%
\providecommand \@@href[1]{\endgroup#1\@@endlink}%
\providecommand \@sanitize@url [0]{\catcode `\\12\catcode `\$12\catcode
  `\&12\catcode `\#12\catcode `\^12\catcode `\_12\catcode `\%12\relax}%
\providecommand \@@startlink[1]{}%
\providecommand \@@endlink[0]{}%
\providecommand \url  [0]{\begingroup\@sanitize@url \@url }%
\providecommand \@url [1]{\endgroup\@href {#1}{\urlprefix }}%
\providecommand \urlprefix  [0]{URL }%
\providecommand \Eprint [0]{\href }%
\providecommand \doibase [0]{http://dx.doi.org/}%
\providecommand \selectlanguage [0]{\@gobble}%
\providecommand \bibinfo  [0]{\@secondoftwo}%
\providecommand \bibfield  [0]{\@secondoftwo}%
\providecommand \translation [1]{[#1]}%
\providecommand \BibitemOpen [0]{}%
\providecommand \bibitemStop [0]{}%
\providecommand \bibitemNoStop [0]{.\EOS\space}%
\providecommand \EOS [0]{\spacefactor3000\relax}%
\providecommand \BibitemShut  [1]{\csname bibitem#1\endcsname}%
\let\auto@bib@innerbib\@empty
%</preamble>
\bibitem [{\citenamefont {Giorgini}\ \emph {et~al.}(2008)\citenamefont
  {Giorgini}, \citenamefont {Pitaevskii},\ and\ \citenamefont
  {Stringari}}]{Giorgini2008}%
  \BibitemOpen
  \bibfield  {author} {\bibinfo {author} {\bibfnamefont {Stefano}\ \bibnamefont
  {Giorgini}}, \bibinfo {author} {\bibfnamefont {Lev~P.}\ \bibnamefont
  {Pitaevskii}}, \ and\ \bibinfo {author} {\bibfnamefont {Sandro}\ \bibnamefont
  {Stringari}},\ }\bibfield  {title} {\enquote {\bibinfo {title} {Theory of
  ultracold atomic {F}ermi gases},}\ }\href {\doibase
  10.1103/RevModPhys.80.1215} {\bibfield  {journal} {\bibinfo  {journal} {Rev.
  Mod. Phys.}\ }\textbf {\bibinfo {volume} {80}},\ \bibinfo {pages}
  {1215--1274} (\bibinfo {year} {2008})}\BibitemShut {NoStop}%
\bibitem [{\citenamefont {Bloch}\ \emph {et~al.}(2008)\citenamefont {Bloch},
  \citenamefont {Dalibard},\ and\ \citenamefont {Zwerger}}]{Bloch2008}%
  \BibitemOpen
  \bibfield  {author} {\bibinfo {author} {\bibfnamefont {Immanuel}\
  \bibnamefont {Bloch}}, \bibinfo {author} {\bibfnamefont {Jean}\ \bibnamefont
  {Dalibard}}, \ and\ \bibinfo {author} {\bibfnamefont {Wilhelm}\ \bibnamefont
  {Zwerger}},\ }\bibfield  {title} {\enquote {\bibinfo {title} {Many-body
  physics with ultracold gases},}\ }\href {\doibase 10.1103/RevModPhys.80.885}
  {\bibfield  {journal} {\bibinfo  {journal} {Rev. Mod. Phys.}\ }\textbf
  {\bibinfo {volume} {80}},\ \bibinfo {pages} {885--964} (\bibinfo {year}
  {2008})}\BibitemShut {NoStop}%
\bibitem [{\citenamefont {Randeria}\ and\ \citenamefont
  {Taylor}(2014)}]{Randeria2014}%
  \BibitemOpen
  \bibfield  {author} {\bibinfo {author} {\bibfnamefont {Mohit}\ \bibnamefont
  {Randeria}}\ and\ \bibinfo {author} {\bibfnamefont {Edward}\ \bibnamefont
  {Taylor}},\ }\bibfield  {title} {\enquote {\bibinfo {title} {Crossover from
  {B}ardeen-{C}ooper-{S}chrieffer to {B}ose-{E}instein condensation and the
  unitary {F}ermi gas},}\ }\href@noop {} {\bibfield  {journal} {\bibinfo
  {journal} {Annual Review of Condensed Matter Physics}\ }\textbf {\bibinfo
  {volume} {5}},\ \bibinfo {pages} {209--232} (\bibinfo {year}
  {2014})}\BibitemShut {NoStop}%
\bibitem [{\citenamefont {Nascimb\`ene}\ \emph {et~al.}(2010)\citenamefont
  {Nascimb\`ene}, \citenamefont {Navon}, \citenamefont {Jiang}, \citenamefont
  {Chevy},\ and\ \citenamefont {Salomon}}]{Nascimbene2010}%
  \BibitemOpen
  \bibfield  {author} {\bibinfo {author} {\bibfnamefont {S.}~\bibnamefont
  {Nascimb\`ene}}, \bibinfo {author} {\bibfnamefont {N.}~\bibnamefont {Navon}},
  \bibinfo {author} {\bibfnamefont {K.~J.}\ \bibnamefont {Jiang}}, \bibinfo
  {author} {\bibfnamefont {F.}~\bibnamefont {Chevy}}, \ and\ \bibinfo {author}
  {\bibfnamefont {C.}~\bibnamefont {Salomon}},\ }\bibfield  {title} {\enquote
  {\bibinfo {title} {Exploring the thermodynamics of a universal {F}ermi
  gas},}\ }\href {http://dx.doi.org/10.1038/nature08814} {\bibfield  {journal}
  {\bibinfo  {journal} {Nature}\ }\textbf {\bibinfo {volume} {463}},\ \bibinfo
  {pages} {1057--1060} (\bibinfo {year} {2010})}\BibitemShut {NoStop}%
\bibitem [{\citenamefont {Horikoshi}\ \emph {et~al.}(2010)\citenamefont
  {Horikoshi}, \citenamefont {Nakajima}, \citenamefont {Ueda},\ and\
  \citenamefont {Mukaiyama}}]{Horikoshi2010}%
  \BibitemOpen
  \bibfield  {author} {\bibinfo {author} {\bibfnamefont {Munekazu}\
  \bibnamefont {Horikoshi}}, \bibinfo {author} {\bibfnamefont {Shuta}\
  \bibnamefont {Nakajima}}, \bibinfo {author} {\bibfnamefont {Masahito}\
  \bibnamefont {Ueda}}, \ and\ \bibinfo {author} {\bibfnamefont {Takashi}\
  \bibnamefont {Mukaiyama}},\ }\bibfield  {title} {\enquote {\bibinfo {title}
  {Measurement of universal thermodynamic functions for a unitary fermi gas},}\
  }\href {\doibase 10.1126/science.1183012} {\bibfield  {journal} {\bibinfo
  {journal} {Science}\ }\textbf {\bibinfo {volume} {327}},\ \bibinfo {pages}
  {442--445} (\bibinfo {year} {2010})}\BibitemShut {NoStop}%
\bibitem [{\citenamefont {Ku}\ \emph {et~al.}(2012)\citenamefont {Ku},
  \citenamefont {Sommer}, \citenamefont {Cheuk},\ and\ \citenamefont
  {Zwierlein}}]{Ku2012}%
  \BibitemOpen
  \bibfield  {author} {\bibinfo {author} {\bibfnamefont {Mark J.~H.}\
  \bibnamefont {Ku}}, \bibinfo {author} {\bibfnamefont {Ariel~T.}\ \bibnamefont
  {Sommer}}, \bibinfo {author} {\bibfnamefont {Lawrence~W.}\ \bibnamefont
  {Cheuk}}, \ and\ \bibinfo {author} {\bibfnamefont {Martin~W.}\ \bibnamefont
  {Zwierlein}},\ }\bibfield  {title} {\enquote {\bibinfo {title} {Revealing the
  superfluid lambda transition in the universal thermodynamics of a unitary
  {F}ermi gas},}\ }\href
  {http://www.sciencemag.org/content/335/6068/563.abstract} {\bibfield
  {journal} {\bibinfo  {journal} {Science}\ }\textbf {\bibinfo {volume}
  {335}},\ \bibinfo {pages} {563--567} (\bibinfo {year} {2012})}\BibitemShut
  {NoStop}%
\bibitem [{\citenamefont {Horikoshi}\ \emph {et~al.}(2017)\citenamefont
  {Horikoshi}, \citenamefont {Koashi}, \citenamefont {Tajima}, \citenamefont
  {Ohashi},\ and\ \citenamefont {Kuwata-Gonokami}}]{Horikoshi2017}%
  \BibitemOpen
  \bibfield  {author} {\bibinfo {author} {\bibfnamefont {Munekazu}\
  \bibnamefont {Horikoshi}}, \bibinfo {author} {\bibfnamefont {Masato}\
  \bibnamefont {Koashi}}, \bibinfo {author} {\bibfnamefont {Hiroyuki}\
  \bibnamefont {Tajima}}, \bibinfo {author} {\bibfnamefont {Yoji}\ \bibnamefont
  {Ohashi}}, \ and\ \bibinfo {author} {\bibfnamefont {Makoto}\ \bibnamefont
  {Kuwata-Gonokami}},\ }\bibfield  {title} {\enquote {\bibinfo {title}
  {Ground-state thermodynamic quantities of homogeneous spin-$1/2$ fermions
  from the bcs region to the unitarity limit},}\ }\href {\doibase
  10.1103/PhysRevX.7.041004} {\bibfield  {journal} {\bibinfo  {journal} {Phys.
  Rev. X}\ }\textbf {\bibinfo {volume} {7}},\ \bibinfo {pages} {041004}
  (\bibinfo {year} {2017})}\BibitemShut {NoStop}%
\bibitem [{\citenamefont {Ho}(2004)}]{Ho2004}%
  \BibitemOpen
  \bibfield  {author} {\bibinfo {author} {\bibfnamefont {Tin-Lun}\ \bibnamefont
  {Ho}},\ }\bibfield  {title} {\enquote {\bibinfo {title} {Universal
  thermodynamics of degenerate quantum gases in the unitarity limit},}\ }\href
  {\doibase 10.1103/PhysRevLett.92.090402} {\bibfield  {journal} {\bibinfo
  {journal} {Phys. Rev. Lett.}\ }\textbf {\bibinfo {volume} {92}},\ \bibinfo
  {pages} {090402} (\bibinfo {year} {2004})}\BibitemShut {NoStop}%
\bibitem [{\citenamefont {Hu}\ \emph {et~al.}(2007)\citenamefont {Hu},
  \citenamefont {Drummond},\ and\ \citenamefont {Liu}}]{Hu2007}%
  \BibitemOpen
  \bibfield  {author} {\bibinfo {author} {\bibfnamefont {Hui}\ \bibnamefont
  {Hu}}, \bibinfo {author} {\bibfnamefont {Peter~D.}\ \bibnamefont {Drummond}},
  \ and\ \bibinfo {author} {\bibfnamefont {Xia-Ji}\ \bibnamefont {Liu}},\
  }\bibfield  {title} {\enquote {\bibinfo {title} {Universal thermodynamics of
  strongly interacting {F}ermi gases},}\ }\href
  {http://dx.doi.org/10.1038/nphys598} {\bibfield  {journal} {\bibinfo
  {journal} {Nat Phys}\ }\textbf {\bibinfo {volume} {3}},\ \bibinfo {pages}
  {469--472} (\bibinfo {year} {2007})}\BibitemShut {NoStop}%
\bibitem [{\citenamefont {Lee}\ and\ \citenamefont
  {Sch\"afer}(2006)}]{Lee2006}%
  \BibitemOpen
  \bibfield  {author} {\bibinfo {author} {\bibfnamefont {Dean}\ \bibnamefont
  {Lee}}\ and\ \bibinfo {author} {\bibfnamefont {Thomas}\ \bibnamefont
  {Sch\"afer}},\ }\bibfield  {title} {\enquote {\bibinfo {title} {Cold dilute
  neutron matter on the lattice. {I}. {L}attice virial coefficients and large
  scattering lengths},}\ }\href {\doibase 10.1103/PhysRevC.73.015201}
  {\bibfield  {journal} {\bibinfo  {journal} {Phys. Rev. C}\ }\textbf {\bibinfo
  {volume} {73}},\ \bibinfo {pages} {015201} (\bibinfo {year}
  {2006})}\BibitemShut {NoStop}%
\bibitem [{\citenamefont {Hen}\ and\ \citenamefont {{\it et
  al}}(2014)}]{Hen2014}%
  \BibitemOpen
  \bibfield  {author} {\bibinfo {author} {\bibfnamefont {O.}~\bibnamefont
  {Hen}}\ and\ \bibinfo {author} {\bibnamefont {{\it et al}}},\ }\bibfield
  {title} {\enquote {\bibinfo {title} {Momentum sharing in imbalanced {F}ermi
  systems},}\ }\href {https://science.sciencemag.org/content/346/6209/614}
  {\bibfield  {journal} {\bibinfo  {journal} {Science}\ }\textbf {\bibinfo
  {volume} {346}},\ \bibinfo {pages} {614--617} (\bibinfo {year}
  {2014})}\BibitemShut {NoStop}%
\bibitem [{\citenamefont {van Wyk}\ \emph {et~al.}(2018)\citenamefont {van
  Wyk}, \citenamefont {Tajima}, \citenamefont {Inotani}, \citenamefont
  {Ohnishi},\ and\ \citenamefont {Ohashi}}]{vanWyk2018}%
  \BibitemOpen
  \bibfield  {author} {\bibinfo {author} {\bibfnamefont {Pieter}\ \bibnamefont
  {van Wyk}}, \bibinfo {author} {\bibfnamefont {Hiroyuki}\ \bibnamefont
  {Tajima}}, \bibinfo {author} {\bibfnamefont {Daisuke}\ \bibnamefont
  {Inotani}}, \bibinfo {author} {\bibfnamefont {Akira}\ \bibnamefont
  {Ohnishi}}, \ and\ \bibinfo {author} {\bibfnamefont {Yoji}\ \bibnamefont
  {Ohashi}},\ }\bibfield  {title} {\enquote {\bibinfo {title} {Superfluid
  {F}ermi atomic gas as a quantum simulator for the study of the neutron-star
  equation of state in the low-density region},}\ }\href {\doibase
  10.1103/PhysRevA.97.013601} {\bibfield  {journal} {\bibinfo  {journal} {Phys.
  Rev. A}\ }\textbf {\bibinfo {volume} {97}},\ \bibinfo {pages} {013601}
  (\bibinfo {year} {2018})}\BibitemShut {NoStop}%
\bibitem [{\citenamefont {Ohashi}\ \emph {et~al.}(2020)\citenamefont {Ohashi},
  \citenamefont {Tajima},\ and\ \citenamefont {{van Wyk}}}]{Ohashi2020}%
  \BibitemOpen
  \bibfield  {author} {\bibinfo {author} {\bibfnamefont {Y.}~\bibnamefont
  {Ohashi}}, \bibinfo {author} {\bibfnamefont {H.}~\bibnamefont {Tajima}}, \
  and\ \bibinfo {author} {\bibfnamefont {P.}~\bibnamefont {{van Wyk}}},\
  }\bibfield  {title} {\enquote {\bibinfo {title} {{BCS}-{BEC} crossover in
  cold atomic and in nuclear systems},}\ }\href {\doibase
  https://doi.org/10.1016/j.ppnp.2019.103739} {\bibfield  {journal} {\bibinfo
  {journal} {Progress in Particle and Nuclear Physics}\ }\textbf {\bibinfo
  {volume} {111}},\ \bibinfo {pages} {103739} (\bibinfo {year}
  {2020})}\BibitemShut {NoStop}%
\bibitem [{\citenamefont {Loktev}\ \emph {et~al.}(2001)\citenamefont {Loktev},
  \citenamefont {Quick},\ and\ \citenamefont {Sharapov}}]{loktev2001}%
  \BibitemOpen
  \bibfield  {author} {\bibinfo {author} {\bibfnamefont {Vadim~M}\ \bibnamefont
  {Loktev}}, \bibinfo {author} {\bibfnamefont {Rachel~M}\ \bibnamefont
  {Quick}}, \ and\ \bibinfo {author} {\bibfnamefont {Sergei~G}\ \bibnamefont
  {Sharapov}},\ }\bibfield  {title} {\enquote {\bibinfo {title} {Phase
  fluctuations and pseudogap phenomena},}\ }\href@noop {} {\bibfield  {journal}
  {\bibinfo  {journal} {Physics Reports}\ }\textbf {\bibinfo {volume} {349}},\
  \bibinfo {pages} {1--123} (\bibinfo {year} {2001})}\BibitemShut {NoStop}%
\bibitem [{\citenamefont {Stajic}(2017)}]{Stajic2017}%
  \BibitemOpen
  \bibfield  {author} {\bibinfo {author} {\bibfnamefont {Jelena}\ \bibnamefont
  {Stajic}},\ }\bibfield  {title} {\enquote {\bibinfo {title} {Making sense of
  the cuprate pseudogap},}\ }\href {\doibase 10.1126/science.357.6351.561-e}
  {\bibfield  {journal} {\bibinfo  {journal} {Science}\ }\textbf {\bibinfo
  {volume} {357}},\ \bibinfo {pages} {561--562} (\bibinfo {year}
  {2017})}\BibitemShut {NoStop}%
\bibitem [{\citenamefont {Chin}\ \emph {et~al.}(2010)\citenamefont {Chin},
  \citenamefont {Grimm}, \citenamefont {Julienne},\ and\ \citenamefont
  {Tiesinga}}]{Chin2010}%
  \BibitemOpen
  \bibfield  {author} {\bibinfo {author} {\bibfnamefont {Cheng}\ \bibnamefont
  {Chin}}, \bibinfo {author} {\bibfnamefont {Rudolf}\ \bibnamefont {Grimm}},
  \bibinfo {author} {\bibfnamefont {Paul}\ \bibnamefont {Julienne}}, \ and\
  \bibinfo {author} {\bibfnamefont {Eite}\ \bibnamefont {Tiesinga}},\
  }\bibfield  {title} {\enquote {\bibinfo {title} {Feshbach resonances in
  ultracold gases},}\ }\href
  {http://link.aps.org/doi/10.1103/RevModPhys.82.1225} {\bibfield  {journal}
  {\bibinfo  {journal} {Rev. Mod. Phys.}\ }\textbf {\bibinfo {volume} {82}},\
  \bibinfo {pages} {1225--1286} (\bibinfo {year} {2010})}\BibitemShut {NoStop}%
\bibitem [{\citenamefont {Nikolic}\ and\ \citenamefont
  {Sachdev}(2007)}]{Nikolic2007}%
  \BibitemOpen
  \bibfield  {author} {\bibinfo {author} {\bibfnamefont {Predrag}\ \bibnamefont
  {Nikolic}}\ and\ \bibinfo {author} {\bibfnamefont {Subir}\ \bibnamefont
  {Sachdev}},\ }\bibfield  {title} {\enquote {\bibinfo {title}
  {Renormalization-group fixed points, universal phase diagram, and $1/n$
  expansion for quantum liquids with interactions near the unitarity limit},}\
  }\href {\doibase 10.1103/PhysRevA.75.033608} {\bibfield  {journal} {\bibinfo
  {journal} {Phys. Rev. A}\ }\textbf {\bibinfo {volume} {75}},\ \bibinfo
  {pages} {033608} (\bibinfo {year} {2007})}\BibitemShut {NoStop}%
\bibitem [{\citenamefont {Tan}(2008{\natexlab{a}})}]{Tan1}%
  \BibitemOpen
  \bibfield  {author} {\bibinfo {author} {\bibfnamefont {Shina}\ \bibnamefont
  {Tan}},\ }\bibfield  {title} {\enquote {\bibinfo {title} {Generalized virial
  theorem and pressure relation for a strongly correlated {F}ermi gas},}\
  }\href {\doibase http://dx.doi.org/10.1016/j.aop.2008.03.003} {\bibfield
  {journal} {\bibinfo  {journal} {Annals of Physics}\ }\textbf {\bibinfo
  {volume} {323}},\ \bibinfo {pages} {2987 -- 2990} (\bibinfo {year}
  {2008}{\natexlab{a}})}\BibitemShut {NoStop}%
\bibitem [{\citenamefont {Tan}(2008{\natexlab{b}})}]{Tan2}%
  \BibitemOpen
  \bibfield  {author} {\bibinfo {author} {\bibfnamefont {Shina}\ \bibnamefont
  {Tan}},\ }\bibfield  {title} {\enquote {\bibinfo {title} {Large momentum part
  of a strongly correlated {F}ermi gas},}\ }\href {\doibase
  http://dx.doi.org/10.1016/j.aop.2008.03.005} {\bibfield  {journal} {\bibinfo
  {journal} {Annals of Physics}\ }\textbf {\bibinfo {volume} {323}},\ \bibinfo
  {pages} {2971 -- 2986} (\bibinfo {year} {2008}{\natexlab{b}})}\BibitemShut
  {NoStop}%
\bibitem [{\citenamefont {Tan}(2008{\natexlab{c}})}]{Tan3}%
  \BibitemOpen
  \bibfield  {author} {\bibinfo {author} {\bibfnamefont {Shina}\ \bibnamefont
  {Tan}},\ }\bibfield  {title} {\enquote {\bibinfo {title} {Energetics of a
  strongly correlated {F}ermi gas},}\ }\href {\doibase
  http://dx.doi.org/10.1016/j.aop.2008.03.004} {\bibfield  {journal} {\bibinfo
  {journal} {Annals of Physics}\ }\textbf {\bibinfo {volume} {323}},\ \bibinfo
  {pages} {2952 -- 2970} (\bibinfo {year} {2008}{\natexlab{c}})}\BibitemShut
  {NoStop}%
\bibitem [{\citenamefont {Braaten}\ and\ \citenamefont
  {Platter}(2008)}]{Braaten2008}%
  \BibitemOpen
  \bibfield  {author} {\bibinfo {author} {\bibfnamefont {Eric}\ \bibnamefont
  {Braaten}}\ and\ \bibinfo {author} {\bibfnamefont {Lucas}\ \bibnamefont
  {Platter}},\ }\bibfield  {title} {\enquote {\bibinfo {title} {Exact relations
  for a strongly interacting {F}ermi gas from the operator product
  expansion},}\ }\href {\doibase 10.1103/PhysRevLett.100.205301} {\bibfield
  {journal} {\bibinfo  {journal} {Phys. Rev. Lett.}\ }\textbf {\bibinfo
  {volume} {100}},\ \bibinfo {pages} {205301} (\bibinfo {year}
  {2008})}\BibitemShut {NoStop}%
\bibitem [{\citenamefont {Zhang}\ and\ \citenamefont
  {Leggett}(2009)}]{Zhang2009}%
  \BibitemOpen
  \bibfield  {author} {\bibinfo {author} {\bibfnamefont {Shizhong}\
  \bibnamefont {Zhang}}\ and\ \bibinfo {author} {\bibfnamefont {Anthony~J.}\
  \bibnamefont {Leggett}},\ }\bibfield  {title} {\enquote {\bibinfo {title}
  {Universal properties of the ultracold fermi gas},}\ }\href {\doibase
  10.1103/PhysRevA.79.023601} {\bibfield  {journal} {\bibinfo  {journal} {Phys.
  Rev. A}\ }\textbf {\bibinfo {volume} {79}},\ \bibinfo {pages} {023601}
  (\bibinfo {year} {2009})}\BibitemShut {NoStop}%
\bibitem [{\citenamefont {Stewart}\ \emph {et~al.}(2010)\citenamefont
  {Stewart}, \citenamefont {Gaebler}, \citenamefont {Drake},\ and\
  \citenamefont {Jin}}]{Stewart2010}%
  \BibitemOpen
  \bibfield  {author} {\bibinfo {author} {\bibfnamefont {J.~T.}\ \bibnamefont
  {Stewart}}, \bibinfo {author} {\bibfnamefont {J.~P.}\ \bibnamefont
  {Gaebler}}, \bibinfo {author} {\bibfnamefont {T.~E.}\ \bibnamefont {Drake}},
  \ and\ \bibinfo {author} {\bibfnamefont {D.~S.}\ \bibnamefont {Jin}},\
  }\bibfield  {title} {\enquote {\bibinfo {title} {Verification of universal
  relations in a strongly interacting fermi gas},}\ }\href {\doibase
  10.1103/PhysRevLett.104.235301} {\bibfield  {journal} {\bibinfo  {journal}
  {Phys. Rev. Lett.}\ }\textbf {\bibinfo {volume} {104}},\ \bibinfo {pages}
  {235301} (\bibinfo {year} {2010})}\BibitemShut {NoStop}%
\bibitem [{\citenamefont {Kuhnle}\ \emph {et~al.}(2010)\citenamefont {Kuhnle},
  \citenamefont {Hu}, \citenamefont {Liu}, \citenamefont {Dyke}, \citenamefont
  {Mark}, \citenamefont {Drummond}, \citenamefont {Hannaford},\ and\
  \citenamefont {Vale}}]{Kuhnle2010}%
  \BibitemOpen
  \bibfield  {author} {\bibinfo {author} {\bibfnamefont {E.~D.}\ \bibnamefont
  {Kuhnle}}, \bibinfo {author} {\bibfnamefont {H.}~\bibnamefont {Hu}}, \bibinfo
  {author} {\bibfnamefont {X.-J.}\ \bibnamefont {Liu}}, \bibinfo {author}
  {\bibfnamefont {P.}~\bibnamefont {Dyke}}, \bibinfo {author} {\bibfnamefont
  {M.}~\bibnamefont {Mark}}, \bibinfo {author} {\bibfnamefont {P.~D.}\
  \bibnamefont {Drummond}}, \bibinfo {author} {\bibfnamefont {P.}~\bibnamefont
  {Hannaford}}, \ and\ \bibinfo {author} {\bibfnamefont {C.~J.}\ \bibnamefont
  {Vale}},\ }\bibfield  {title} {\enquote {\bibinfo {title} {Universal behavior
  of pair correlations in a strongly interacting fermi gas},}\ }\href {\doibase
  10.1103/PhysRevLett.105.070402} {\bibfield  {journal} {\bibinfo  {journal}
  {Phys. Rev. Lett.}\ }\textbf {\bibinfo {volume} {105}},\ \bibinfo {pages}
  {070402} (\bibinfo {year} {2010})}\BibitemShut {NoStop}%
\bibitem [{\citenamefont {Werner}\ and\ \citenamefont
  {Castin}(2012)}]{Werner2012}%
  \BibitemOpen
  \bibfield  {author} {\bibinfo {author} {\bibfnamefont {F\'elix}\ \bibnamefont
  {Werner}}\ and\ \bibinfo {author} {\bibfnamefont {Yvan}\ \bibnamefont
  {Castin}},\ }\bibfield  {title} {\enquote {\bibinfo {title} {General
  relations for quantum gases in two and three dimensions: {T}wo-component
  fermions},}\ }\href {\doibase 10.1103/PhysRevA.86.013626} {\bibfield
  {journal} {\bibinfo  {journal} {Phys. Rev. A}\ }\textbf {\bibinfo {volume}
  {86}},\ \bibinfo {pages} {013626} (\bibinfo {year} {2012})}\BibitemShut
  {NoStop}%
\bibitem [{\citenamefont {Braaten}(2012)}]{Braaten2012}%
  \BibitemOpen
  \bibfield  {author} {\bibinfo {author} {\bibfnamefont {Eric}\ \bibnamefont
  {Braaten}},\ }\enquote {\bibinfo {title} {Universal relations for fermions
  with large scattering length},}\ in\ \href {\doibase
  10.1007/978-3-642-21978-8_6} {\emph {\bibinfo {booktitle} {The BCS-BEC
  crossover and the unitary Fermi gas}}},\ \bibinfo {editor} {edited by\
  \bibinfo {editor} {\bibfnamefont {Wilhelm}\ \bibnamefont {Zwerger}}}\
  (\bibinfo  {publisher} {Springer Berlin Heidelberg},\ \bibinfo {address}
  {Berlin, Heidelberg},\ \bibinfo {year} {2012})\ pp.\ \bibinfo {pages}
  {193--231}\BibitemShut {NoStop}%
\bibitem [{\citenamefont {Forbes}\ \emph {et~al.}(2011)\citenamefont {Forbes},
  \citenamefont {Gandolfi},\ and\ \citenamefont {Gezerlis}}]{Forbes2011}%
  \BibitemOpen
  \bibfield  {author} {\bibinfo {author} {\bibfnamefont {Michael~McNeil}\
  \bibnamefont {Forbes}}, \bibinfo {author} {\bibfnamefont {Stefano}\
  \bibnamefont {Gandolfi}}, \ and\ \bibinfo {author} {\bibfnamefont
  {Alexandros}\ \bibnamefont {Gezerlis}},\ }\bibfield  {title} {\enquote
  {\bibinfo {title} {Resonantly interacting fermions in a box},}\ }\href
  {\doibase 10.1103/PhysRevLett.106.235303} {\bibfield  {journal} {\bibinfo
  {journal} {Phys. Rev. Lett.}\ }\textbf {\bibinfo {volume} {106}},\ \bibinfo
  {pages} {235303} (\bibinfo {year} {2011})}\BibitemShut {NoStop}%
\bibitem [{\citenamefont {Gandolfi}\ \emph {et~al.}(2011)\citenamefont
  {Gandolfi}, \citenamefont {Schmidt},\ and\ \citenamefont
  {Carlson}}]{Carlson2011}%
  \BibitemOpen
  \bibfield  {author} {\bibinfo {author} {\bibfnamefont {S.}~\bibnamefont
  {Gandolfi}}, \bibinfo {author} {\bibfnamefont {K.~E.}\ \bibnamefont
  {Schmidt}}, \ and\ \bibinfo {author} {\bibfnamefont {J.}~\bibnamefont
  {Carlson}},\ }\bibfield  {title} {\enquote {\bibinfo {title} {{BEC}-{BCS}
  crossover and universal relations in unitary {F}ermi gases},}\ }\href
  {\doibase 10.1103/PhysRevA.83.041601} {\bibfield  {journal} {\bibinfo
  {journal} {Phys. Rev. A}\ }\textbf {\bibinfo {volume} {83}},\ \bibinfo
  {pages} {041601} (\bibinfo {year} {2011})}\BibitemShut {NoStop}%
\bibitem [{\citenamefont {Drut}(2012)}]{Drut2012}%
  \BibitemOpen
  \bibfield  {author} {\bibinfo {author} {\bibfnamefont {Joaqu\'{\i}n~E.}\
  \bibnamefont {Drut}},\ }\bibfield  {title} {\enquote {\bibinfo {title}
  {Improved lattice operators for nonrelativistic fermions},}\ }\href {\doibase
  10.1103/PhysRevA.86.013604} {\bibfield  {journal} {\bibinfo  {journal} {Phys.
  Rev. A}\ }\textbf {\bibinfo {volume} {86}},\ \bibinfo {pages} {013604}
  (\bibinfo {year} {2012})}\BibitemShut {NoStop}%
\bibitem [{\citenamefont {Goulko}\ and\ \citenamefont
  {Wingate}(2016)}]{Goulko2016QMC}%
  \BibitemOpen
  \bibfield  {author} {\bibinfo {author} {\bibfnamefont {Olga}\ \bibnamefont
  {Goulko}}\ and\ \bibinfo {author} {\bibfnamefont {Matthew}\ \bibnamefont
  {Wingate}},\ }\bibfield  {title} {\enquote {\bibinfo {title} {Numerical study
  of the unitary fermi gas across the superfluid transition},}\ }\href
  {\doibase 10.1103/PhysRevA.93.053604} {\bibfield  {journal} {\bibinfo
  {journal} {Phys. Rev. A}\ }\textbf {\bibinfo {volume} {93}},\ \bibinfo
  {pages} {053604} (\bibinfo {year} {2016})}\BibitemShut {NoStop}%
\bibitem [{\citenamefont {Jensen}\ \emph {et~al.}(2020)\citenamefont {Jensen},
  \citenamefont {Gilbreth},\ and\ \citenamefont {Alhassid}}]{Jensen2020}%
  \BibitemOpen
  \bibfield  {author} {\bibinfo {author} {\bibfnamefont {S.}~\bibnamefont
  {Jensen}}, \bibinfo {author} {\bibfnamefont {C.~N.}\ \bibnamefont
  {Gilbreth}}, \ and\ \bibinfo {author} {\bibfnamefont {Y.}~\bibnamefont
  {Alhassid}},\ }\bibfield  {title} {\enquote {\bibinfo {title} {Pairing
  correlations across the superfluid phase transition in the unitary {F}ermi
  gas},}\ }\href {\doibase 10.1103/PhysRevLett.124.090604} {\bibfield
  {journal} {\bibinfo  {journal} {Phys. Rev. Lett.}\ }\textbf {\bibinfo
  {volume} {124}},\ \bibinfo {pages} {090604} (\bibinfo {year}
  {2020})}\BibitemShut {NoStop}%
\bibitem [{\citenamefont {Richie-Halford}\ \emph {et~al.}(2020)\citenamefont
  {Richie-Halford}, \citenamefont {Drut},\ and\ \citenamefont
  {Bulgac}}]{Halford2020}%
  \BibitemOpen
  \bibfield  {author} {\bibinfo {author} {\bibfnamefont {Adam}\ \bibnamefont
  {Richie-Halford}}, \bibinfo {author} {\bibfnamefont {Joaqu\'{\i}n~E.}\
  \bibnamefont {Drut}}, \ and\ \bibinfo {author} {\bibfnamefont {Aurel}\
  \bibnamefont {Bulgac}},\ }\bibfield  {title} {\enquote {\bibinfo {title}
  {Emergence of a pseudogap in the {BCS-BEC} crossover},}\ }\href {\doibase
  10.1103/PhysRevLett.125.060403} {\bibfield  {journal} {\bibinfo  {journal}
  {Phys. Rev. Lett.}\ }\textbf {\bibinfo {volume} {125}},\ \bibinfo {pages}
  {060403} (\bibinfo {year} {2020})}\BibitemShut {NoStop}%
\bibitem [{\citenamefont {Van~Houcke}\ \emph {et~al.}(2012)\citenamefont
  {Van~Houcke}, \citenamefont {Werner}, \citenamefont {Kozik}, \citenamefont
  {Prokof'ev}, \citenamefont {Svistunov}, \citenamefont {Ku}, \citenamefont
  {Sommer}, \citenamefont {Cheuk}, \citenamefont {Schirotzek},\ and\
  \citenamefont {Zwierlein}}]{VanHoucke2012}%
  \BibitemOpen
  \bibfield  {author} {\bibinfo {author} {\bibfnamefont {K.}~\bibnamefont
  {Van~Houcke}}, \bibinfo {author} {\bibfnamefont {F.}~\bibnamefont {Werner}},
  \bibinfo {author} {\bibfnamefont {E.}~\bibnamefont {Kozik}}, \bibinfo
  {author} {\bibfnamefont {N.}~\bibnamefont {Prokof'ev}}, \bibinfo {author}
  {\bibfnamefont {B.}~\bibnamefont {Svistunov}}, \bibinfo {author}
  {\bibfnamefont {M.~J.~H.}\ \bibnamefont {Ku}}, \bibinfo {author}
  {\bibfnamefont {A.~T.}\ \bibnamefont {Sommer}}, \bibinfo {author}
  {\bibfnamefont {L.~W.}\ \bibnamefont {Cheuk}}, \bibinfo {author}
  {\bibfnamefont {A.}~\bibnamefont {Schirotzek}}, \ and\ \bibinfo {author}
  {\bibfnamefont {M.~W.}\ \bibnamefont {Zwierlein}},\ }\bibfield  {title}
  {\enquote {\bibinfo {title} {Feynman diagrams versus {F}ermi-gas {F}eynman
  emulator},}\ }\href {http://dx.doi.org/10.1038/nphys2273} {\bibfield
  {journal} {\bibinfo  {journal} {Nat Phys}\ }\textbf {\bibinfo {volume} {8}},\
  \bibinfo {pages} {366--370} (\bibinfo {year} {2012})}\BibitemShut {NoStop}%
\bibitem [{\citenamefont {Drut}\ \emph {et~al.}(2012)\citenamefont {Drut},
  \citenamefont {L\"ahde}, \citenamefont {Wlaz\l{}owski},\ and\ \citenamefont
  {Magierski}}]{Drut2012b}%
  \BibitemOpen
  \bibfield  {author} {\bibinfo {author} {\bibfnamefont {Joaqu\'{\i}n~E.}\
  \bibnamefont {Drut}}, \bibinfo {author} {\bibfnamefont {Timo~A.}\
  \bibnamefont {L\"ahde}}, \bibinfo {author} {\bibfnamefont {Gabriel}\
  \bibnamefont {Wlaz\l{}owski}}, \ and\ \bibinfo {author} {\bibfnamefont
  {Piotr}\ \bibnamefont {Magierski}},\ }\bibfield  {title} {\enquote {\bibinfo
  {title} {Equation of state of the unitary {F}ermi gas: An update on lattice
  calculations},}\ }\href {\doibase 10.1103/PhysRevA.85.051601} {\bibfield
  {journal} {\bibinfo  {journal} {Phys. Rev. A}\ }\textbf {\bibinfo {volume}
  {85}},\ \bibinfo {pages} {051601} (\bibinfo {year} {2012})}\BibitemShut
  {NoStop}%
\bibitem [{\citenamefont {Rossi}\ \emph {et~al.}(2018)\citenamefont {Rossi},
  \citenamefont {Ohgoe}, \citenamefont {Kozik}, \citenamefont {Prokof'ev},
  \citenamefont {Svistunov}, \citenamefont {Van~Houcke},\ and\ \citenamefont
  {Werner}}]{Rossi2018}%
  \BibitemOpen
  \bibfield  {author} {\bibinfo {author} {\bibfnamefont {R.}~\bibnamefont
  {Rossi}}, \bibinfo {author} {\bibfnamefont {T.}~\bibnamefont {Ohgoe}},
  \bibinfo {author} {\bibfnamefont {E.}~\bibnamefont {Kozik}}, \bibinfo
  {author} {\bibfnamefont {N.}~\bibnamefont {Prokof'ev}}, \bibinfo {author}
  {\bibfnamefont {B.}~\bibnamefont {Svistunov}}, \bibinfo {author}
  {\bibfnamefont {K.}~\bibnamefont {Van~Houcke}}, \ and\ \bibinfo {author}
  {\bibfnamefont {F.}~\bibnamefont {Werner}},\ }\bibfield  {title} {\enquote
  {\bibinfo {title} {Contact and momentum distribution of the unitary {F}ermi
  gas},}\ }\href {\doibase 10.1103/PhysRevLett.121.130406} {\bibfield
  {journal} {\bibinfo  {journal} {Phys. Rev. Lett.}\ }\textbf {\bibinfo
  {volume} {121}},\ \bibinfo {pages} {130406} (\bibinfo {year}
  {2018})}\BibitemShut {NoStop}%
\bibitem [{\citenamefont {Nozieres}\ and\ \citenamefont
  {Schmitt-Rink}(1985)}]{nozieres1985bose}%
  \BibitemOpen
  \bibfield  {author} {\bibinfo {author} {\bibfnamefont {Ph}~\bibnamefont
  {Nozieres}}\ and\ \bibinfo {author} {\bibfnamefont {S}~\bibnamefont
  {Schmitt-Rink}},\ }\bibfield  {title} {\enquote {\bibinfo {title} {Bose
  condensation in an attractive fermion gas: {F}rom weak to strong coupling
  superconductivity},}\ }\href@noop {} {\bibfield  {journal} {\bibinfo
  {journal} {Journal of Low Temperature Physics}\ }\textbf {\bibinfo {volume}
  {59}},\ \bibinfo {pages} {195--211} (\bibinfo {year} {1985})}\BibitemShut
  {NoStop}%
\bibitem [{\citenamefont {De~Melo}\ \emph {et~al.}(1993)\citenamefont
  {De~Melo}, \citenamefont {Randeria},\ and\ \citenamefont
  {Engelbrecht}}]{de1993crossover}%
  \BibitemOpen
  \bibfield  {author} {\bibinfo {author} {\bibfnamefont {CAR~S{\'a}}\
  \bibnamefont {De~Melo}}, \bibinfo {author} {\bibfnamefont {Mohit}\
  \bibnamefont {Randeria}}, \ and\ \bibinfo {author} {\bibfnamefont {Jan~R}\
  \bibnamefont {Engelbrecht}},\ }\bibfield  {title} {\enquote {\bibinfo {title}
  {Crossover from {{BCS}} to {B}ose superconductivity: {T}ransition temperature
  and time-dependent {G}inzburg-{L}andau theory},}\ }\href@noop {} {\bibfield
  {journal} {\bibinfo  {journal} {Physical review letters}\ }\textbf {\bibinfo
  {volume} {71}},\ \bibinfo {pages} {3202} (\bibinfo {year}
  {1993})}\BibitemShut {NoStop}%
\bibitem [{\citenamefont {Ohashi}\ and\ \citenamefont
  {Griffin}(2003)}]{Ohashi2003}%
  \BibitemOpen
  \bibfield  {author} {\bibinfo {author} {\bibfnamefont {Y.}~\bibnamefont
  {Ohashi}}\ and\ \bibinfo {author} {\bibfnamefont {A.}~\bibnamefont
  {Griffin}},\ }\bibfield  {title} {\enquote {\bibinfo {title} {Superfluidity
  and collective modes in a uniform gas of fermi atoms with a feshbach
  resonance},}\ }\href {\doibase 10.1103/PhysRevA.67.063612} {\bibfield
  {journal} {\bibinfo  {journal} {Phys. Rev. A}\ }\textbf {\bibinfo {volume}
  {67}},\ \bibinfo {pages} {063612} (\bibinfo {year} {2003})}\BibitemShut
  {NoStop}%
\bibitem [{\citenamefont {Pieri}\ \emph {et~al.}(2004)\citenamefont {Pieri},
  \citenamefont {Pisani},\ and\ \citenamefont {Strinati}}]{Pieri2004}%
  \BibitemOpen
  \bibfield  {author} {\bibinfo {author} {\bibfnamefont {P.}~\bibnamefont
  {Pieri}}, \bibinfo {author} {\bibfnamefont {L.}~\bibnamefont {Pisani}}, \
  and\ \bibinfo {author} {\bibfnamefont {G.~C.}\ \bibnamefont {Strinati}},\
  }\bibfield  {title} {\enquote {\bibinfo {title} {{BCS}-{BEC} crossover at
  finite temperature in the broken-symmetry phase},}\ }\href {\doibase
  10.1103/PhysRevB.70.094508} {\bibfield  {journal} {\bibinfo  {journal} {Phys.
  Rev. B}\ }\textbf {\bibinfo {volume} {70}},\ \bibinfo {pages} {094508}
  (\bibinfo {year} {2004})}\BibitemShut {NoStop}%
\bibitem [{\citenamefont {Chen}\ \emph {et~al.}(2005)\citenamefont {Chen},
  \citenamefont {Stajic}, \citenamefont {Tan},\ and\ \citenamefont
  {Levin}}]{Chen2005}%
  \BibitemOpen
  \bibfield  {author} {\bibinfo {author} {\bibfnamefont {Qijin}\ \bibnamefont
  {Chen}}, \bibinfo {author} {\bibfnamefont {Jelena}\ \bibnamefont {Stajic}},
  \bibinfo {author} {\bibfnamefont {Shina}\ \bibnamefont {Tan}}, \ and\
  \bibinfo {author} {\bibfnamefont {K.}~\bibnamefont {Levin}},\ }\bibfield
  {title} {\enquote {\bibinfo {title} {{BCS--BEC} crossover: From high
  temperature superconductors to ultracold superfluids},}\ }\href {\doibase
  http://dx.doi.org/10.1016/j.physrep.2005.02.005} {\bibfield  {journal}
  {\bibinfo  {journal} {Physics Reports}\ }\textbf {\bibinfo {volume} {412}},\
  \bibinfo {pages} {1 -- 88} (\bibinfo {year} {2005})}\BibitemShut {NoStop}%
\bibitem [{\citenamefont {Liu}\ and\ \citenamefont {Hu}(2005)}]{Liu2005PRA}%
  \BibitemOpen
  \bibfield  {author} {\bibinfo {author} {\bibfnamefont {Xia-Ji}\ \bibnamefont
  {Liu}}\ and\ \bibinfo {author} {\bibfnamefont {Hui}\ \bibnamefont {Hu}},\
  }\bibfield  {title} {\enquote {\bibinfo {title} {Self-consistent theory of
  atomic {F}ermi gases with a {F}eshbach resonance at the superfluid
  transition},}\ }\href {\doibase 10.1103/PhysRevA.72.063613} {\bibfield
  {journal} {\bibinfo  {journal} {Phys. Rev. A}\ }\textbf {\bibinfo {volume}
  {72}},\ \bibinfo {pages} {063613} (\bibinfo {year} {2005})}\BibitemShut
  {NoStop}%
\bibitem [{\citenamefont {Taylor}\ \emph {et~al.}(2006)\citenamefont {Taylor},
  \citenamefont {Griffin}, \citenamefont {Fukushima},\ and\ \citenamefont
  {Ohashi}}]{Taylor2006}%
  \BibitemOpen
  \bibfield  {author} {\bibinfo {author} {\bibfnamefont {E.}~\bibnamefont
  {Taylor}}, \bibinfo {author} {\bibfnamefont {A.}~\bibnamefont {Griffin}},
  \bibinfo {author} {\bibfnamefont {N.}~\bibnamefont {Fukushima}}, \ and\
  \bibinfo {author} {\bibfnamefont {Y.}~\bibnamefont {Ohashi}},\ }\bibfield
  {title} {\enquote {\bibinfo {title} {Pairing fluctuations and the superfluid
  density through the {BCS}-{BEC} crossover},}\ }\href {\doibase
  10.1103/PhysRevA.74.063626} {\bibfield  {journal} {\bibinfo  {journal} {Phys.
  Rev. A}\ }\textbf {\bibinfo {volume} {74}},\ \bibinfo {pages} {063626}
  (\bibinfo {year} {2006})}\BibitemShut {NoStop}%
\bibitem [{\citenamefont {Hu}\ \emph {et~al.}(2006)\citenamefont {Hu},
  \citenamefont {Liu},\ and\ \citenamefont {Drummond}}]{Hu2006}%
  \BibitemOpen
  \bibfield  {author} {\bibinfo {author} {\bibfnamefont {H}~\bibnamefont {Hu}},
  \bibinfo {author} {\bibfnamefont {X.-J}\ \bibnamefont {Liu}}, \ and\ \bibinfo
  {author} {\bibfnamefont {P.~D}\ \bibnamefont {Drummond}},\ }\bibfield
  {title} {\enquote {\bibinfo {title} {Equation of state of a superfluid
  {F}ermi gas in the {BCS}-{BEC} crossover},}\ }\href {\doibase
  10.1209/epl/i2006-10023-y} {\bibfield  {journal} {\bibinfo  {journal}
  {Europhysics Letters ({EPL})}\ }\textbf {\bibinfo {volume} {74}},\ \bibinfo
  {pages} {574--580} (\bibinfo {year} {2006})}\BibitemShut {NoStop}%
\bibitem [{\citenamefont {Haussmann}\ \emph {et~al.}(2007)\citenamefont
  {Haussmann}, \citenamefont {Rantner}, \citenamefont {Cerrito},\ and\
  \citenamefont {Zwerger}}]{Haussmann2007}%
  \BibitemOpen
  \bibfield  {author} {\bibinfo {author} {\bibfnamefont {R.}~\bibnamefont
  {Haussmann}}, \bibinfo {author} {\bibfnamefont {W.}~\bibnamefont {Rantner}},
  \bibinfo {author} {\bibfnamefont {S.}~\bibnamefont {Cerrito}}, \ and\
  \bibinfo {author} {\bibfnamefont {W.}~\bibnamefont {Zwerger}},\ }\bibfield
  {title} {\enquote {\bibinfo {title} {Thermodynamics of the {{BCS}-BEC}
  crossover},}\ }\href {\doibase 10.1103/PhysRevA.75.023610} {\bibfield
  {journal} {\bibinfo  {journal} {Phys. Rev. A}\ }\textbf {\bibinfo {volume}
  {75}},\ \bibinfo {pages} {023610} (\bibinfo {year} {2007})}\BibitemShut
  {NoStop}%
\bibitem [{\citenamefont {Diener}\ \emph {et~al.}(2008)\citenamefont {Diener},
  \citenamefont {Sensarma},\ and\ \citenamefont {Randeria}}]{Diener2008}%
  \BibitemOpen
  \bibfield  {author} {\bibinfo {author} {\bibfnamefont {Roberto~B.}\
  \bibnamefont {Diener}}, \bibinfo {author} {\bibfnamefont {Rajdeep}\
  \bibnamefont {Sensarma}}, \ and\ \bibinfo {author} {\bibfnamefont {Mohit}\
  \bibnamefont {Randeria}},\ }\bibfield  {title} {\enquote {\bibinfo {title}
  {Quantum fluctuations in the superfluid state of the {{BCS}-BEC}
  crossover},}\ }\href {\doibase 10.1103/PhysRevA.77.023626} {\bibfield
  {journal} {\bibinfo  {journal} {Phys. Rev. A}\ }\textbf {\bibinfo {volume}
  {77}},\ \bibinfo {pages} {023626} (\bibinfo {year} {2008})}\BibitemShut
  {NoStop}%
\bibitem [{\citenamefont {Hu}\ \emph {et~al.}(2008)\citenamefont {Hu},
  \citenamefont {Liu},\ and\ \citenamefont {Drummond}}]{Huicomparitive08}%
  \BibitemOpen
  \bibfield  {author} {\bibinfo {author} {\bibfnamefont {Hui}\ \bibnamefont
  {Hu}}, \bibinfo {author} {\bibfnamefont {Xia-Ji}\ \bibnamefont {Liu}}, \ and\
  \bibinfo {author} {\bibfnamefont {Peter~D.}\ \bibnamefont {Drummond}},\
  }\bibfield  {title} {\enquote {\bibinfo {title} {Comparative study of
  strong-coupling theories of a trapped {F}ermi gas at unitarity},}\ }\href
  {\doibase 10.1103/PhysRevA.77.061605} {\bibfield  {journal} {\bibinfo
  {journal} {Phys. Rev. A}\ }\textbf {\bibinfo {volume} {77}},\ \bibinfo
  {pages} {061605} (\bibinfo {year} {2008})}\BibitemShut {NoStop}%
\bibitem [{\citenamefont {Watanabe}\ \emph {et~al.}(2010)\citenamefont
  {Watanabe}, \citenamefont {Tsuchiya},\ and\ \citenamefont
  {Ohashi}}]{Watnabe2010}%
  \BibitemOpen
  \bibfield  {author} {\bibinfo {author} {\bibfnamefont {Ryota}\ \bibnamefont
  {Watanabe}}, \bibinfo {author} {\bibfnamefont {Shunji}\ \bibnamefont
  {Tsuchiya}}, \ and\ \bibinfo {author} {\bibfnamefont {Yoji}\ \bibnamefont
  {Ohashi}},\ }\bibfield  {title} {\enquote {\bibinfo {title} {Superfluid
  density of states and pseudogap phenomenon in the {BCS}-{BEC} crossover
  regime of a superfluid {F}ermi gas},}\ }\href {\doibase
  10.1103/PhysRevA.82.043630} {\bibfield  {journal} {\bibinfo  {journal} {Phys.
  Rev. A}\ }\textbf {\bibinfo {volume} {82}},\ \bibinfo {pages} {043630}
  (\bibinfo {year} {2010})}\BibitemShut {NoStop}%
\bibitem [{\citenamefont {He}\ \emph {et~al.}(2015)\citenamefont {He},
  \citenamefont {L\"u}, \citenamefont {Cao}, \citenamefont {Hu},\ and\
  \citenamefont {Liu}}]{He2015}%
  \BibitemOpen
  \bibfield  {author} {\bibinfo {author} {\bibfnamefont {Lianyi}\ \bibnamefont
  {He}}, \bibinfo {author} {\bibfnamefont {Haifeng}\ \bibnamefont {L\"u}},
  \bibinfo {author} {\bibfnamefont {Gaoqing}\ \bibnamefont {Cao}}, \bibinfo
  {author} {\bibfnamefont {Hui}\ \bibnamefont {Hu}}, \ and\ \bibinfo {author}
  {\bibfnamefont {Xia-Ji}\ \bibnamefont {Liu}},\ }\bibfield  {title} {\enquote
  {\bibinfo {title} {Quantum fluctuations in the bcs-{BEC} crossover of
  two-dimensional {F}ermi gases},}\ }\href {\doibase
  10.1103/PhysRevA.92.023620} {\bibfield  {journal} {\bibinfo  {journal} {Phys.
  Rev. A}\ }\textbf {\bibinfo {volume} {92}},\ \bibinfo {pages} {023620}
  (\bibinfo {year} {2015})}\BibitemShut {NoStop}%
\bibitem [{\citenamefont {Mulkerin}\ \emph {et~al.}(2016)\citenamefont
  {Mulkerin}, \citenamefont {Liu},\ and\ \citenamefont {Hu}}]{Mulkerin2016}%
  \BibitemOpen
  \bibfield  {author} {\bibinfo {author} {\bibfnamefont {Brendan~C.}\
  \bibnamefont {Mulkerin}}, \bibinfo {author} {\bibfnamefont {Xia-Ji}\
  \bibnamefont {Liu}}, \ and\ \bibinfo {author} {\bibfnamefont {Hui}\
  \bibnamefont {Hu}},\ }\bibfield  {title} {\enquote {\bibinfo {title} {Beyond
  {G}aussian pair fluctuation theory for strongly interacting {F}ermi gases},}\
  }\href {\doibase 10.1103/PhysRevA.94.013610} {\bibfield  {journal} {\bibinfo
  {journal} {Phys. Rev. A}\ }\textbf {\bibinfo {volume} {94}},\ \bibinfo
  {pages} {013610} (\bibinfo {year} {2016})}\BibitemShut {NoStop}%
\bibitem [{\citenamefont {Tajima}\ \emph {et~al.}(2017)\citenamefont {Tajima},
  \citenamefont {van Wyk}, \citenamefont {Hanai}, \citenamefont {Kagamihara},
  \citenamefont {Inotani}, \citenamefont {Horikoshi},\ and\ \citenamefont
  {Ohashi}}]{Tajima2017}%
  \BibitemOpen
  \bibfield  {author} {\bibinfo {author} {\bibfnamefont {Hiroyuki}\
  \bibnamefont {Tajima}}, \bibinfo {author} {\bibfnamefont {Pieter}\
  \bibnamefont {van Wyk}}, \bibinfo {author} {\bibfnamefont {Ryo}\ \bibnamefont
  {Hanai}}, \bibinfo {author} {\bibfnamefont {Daichi}\ \bibnamefont
  {Kagamihara}}, \bibinfo {author} {\bibfnamefont {Daisuke}\ \bibnamefont
  {Inotani}}, \bibinfo {author} {\bibfnamefont {Munekazu}\ \bibnamefont
  {Horikoshi}}, \ and\ \bibinfo {author} {\bibfnamefont {Yoji}\ \bibnamefont
  {Ohashi}},\ }\bibfield  {title} {\enquote {\bibinfo {title} {Strong-coupling
  corrections to ground-state properties of a superfluid {F}ermi gas},}\ }\href
  {\doibase 10.1103/PhysRevA.95.043625} {\bibfield  {journal} {\bibinfo
  {journal} {Phys. Rev. A}\ }\textbf {\bibinfo {volume} {95}},\ \bibinfo
  {pages} {043625} (\bibinfo {year} {2017})}\BibitemShut {NoStop}%
\bibitem [{\citenamefont {Mulkerin}\ \emph {et~al.}(2017)\citenamefont
  {Mulkerin}, \citenamefont {He}, \citenamefont {Dyke}, \citenamefont {Vale},
  \citenamefont {Liu},\ and\ \citenamefont {Hu}}]{Mulkerin2017}%
  \BibitemOpen
  \bibfield  {author} {\bibinfo {author} {\bibfnamefont {Brendan~C.}\
  \bibnamefont {Mulkerin}}, \bibinfo {author} {\bibfnamefont {Lianyi}\
  \bibnamefont {He}}, \bibinfo {author} {\bibfnamefont {Paul}\ \bibnamefont
  {Dyke}}, \bibinfo {author} {\bibfnamefont {Chris~J.}\ \bibnamefont {Vale}},
  \bibinfo {author} {\bibfnamefont {Xia-Ji}\ \bibnamefont {Liu}}, \ and\
  \bibinfo {author} {\bibfnamefont {Hui}\ \bibnamefont {Hu}},\ }\bibfield
  {title} {\enquote {\bibinfo {title} {Superfluid density and critical velocity
  near the {B}erezinskii-{K}osterlitz-{T}houless transition in a
  two-dimensional strongly interacting {F}ermi gas},}\ }\href {\doibase
  10.1103/PhysRevA.96.053608} {\bibfield  {journal} {\bibinfo  {journal} {Phys.
  Rev. A}\ }\textbf {\bibinfo {volume} {96}},\ \bibinfo {pages} {053608}
  (\bibinfo {year} {2017})}\BibitemShut {NoStop}%
\bibitem [{\citenamefont {Pini}\ \emph {et~al.}(2019)\citenamefont {Pini},
  \citenamefont {Pieri},\ and\ \citenamefont {Strinati}}]{Pini2019}%
  \BibitemOpen
  \bibfield  {author} {\bibinfo {author} {\bibfnamefont {M.}~\bibnamefont
  {Pini}}, \bibinfo {author} {\bibfnamefont {P.}~\bibnamefont {Pieri}}, \ and\
  \bibinfo {author} {\bibfnamefont {G.~Calvanese}\ \bibnamefont {Strinati}},\
  }\bibfield  {title} {\enquote {\bibinfo {title} {Fermi gas throughout the
  bcs-bec crossover: Comparative study of $t$-matrix approaches with various
  degrees of self-consistency},}\ }\href {\doibase 10.1103/PhysRevB.99.094502}
  {\bibfield  {journal} {\bibinfo  {journal} {Phys. Rev. B}\ }\textbf {\bibinfo
  {volume} {99}},\ \bibinfo {pages} {094502} (\bibinfo {year}
  {2019})}\BibitemShut {NoStop}%
\bibitem [{\citenamefont {Nishida}\ and\ \citenamefont
  {Son}(2006)}]{Nishida2006}%
  \BibitemOpen
  \bibfield  {author} {\bibinfo {author} {\bibfnamefont {Yusuke}\ \bibnamefont
  {Nishida}}\ and\ \bibinfo {author} {\bibfnamefont {Dam~Thanh}\ \bibnamefont
  {Son}},\ }\bibfield  {title} {\enquote {\bibinfo {title} {$\epsilon$
  expansion for a fermi gas at infinite scattering length},}\ }\href {\doibase
  10.1103/PhysRevLett.97.050403} {\bibfield  {journal} {\bibinfo  {journal}
  {Phys. Rev. Lett.}\ }\textbf {\bibinfo {volume} {97}},\ \bibinfo {pages}
  {050403} (\bibinfo {year} {2006})}\BibitemShut {NoStop}%
\bibitem [{\citenamefont {Arnold}\ \emph {et~al.}(2007)\citenamefont {Arnold},
  \citenamefont {Drut},\ and\ \citenamefont {Son}}]{Arnold2007}%
  \BibitemOpen
  \bibfield  {author} {\bibinfo {author} {\bibfnamefont {Peter}\ \bibnamefont
  {Arnold}}, \bibinfo {author} {\bibfnamefont {Joaqu\'{\i}n~E.}\ \bibnamefont
  {Drut}}, \ and\ \bibinfo {author} {\bibfnamefont {Dam~Thanh}\ \bibnamefont
  {Son}},\ }\bibfield  {title} {\enquote {\bibinfo {title}
  {Next-to-next-to-leading-order $\epsilon$ expansion for a {F}ermi gas at
  infinite scattering length},}\ }\href {\doibase 10.1103/PhysRevA.75.043605}
  {\bibfield  {journal} {\bibinfo  {journal} {Phys. Rev. A}\ }\textbf {\bibinfo
  {volume} {75}},\ \bibinfo {pages} {043605} (\bibinfo {year}
  {2007})}\BibitemShut {NoStop}%
\bibitem [{\citenamefont {Nishida}\ and\ \citenamefont {Son}(2007)}]{Son2007}%
  \BibitemOpen
  \bibfield  {author} {\bibinfo {author} {\bibfnamefont {Yusuke}\ \bibnamefont
  {Nishida}}\ and\ \bibinfo {author} {\bibfnamefont {Dam~Thanh}\ \bibnamefont
  {Son}},\ }\bibfield  {title} {\enquote {\bibinfo {title} {Fermi gas near
  unitarity around four and two spatial dimensions},}\ }\href {\doibase
  10.1103/PhysRevA.75.063617} {\bibfield  {journal} {\bibinfo  {journal} {Phys.
  Rev. A}\ }\textbf {\bibinfo {volume} {75}},\ \bibinfo {pages} {063617}
  (\bibinfo {year} {2007})}\BibitemShut {NoStop}%
\bibitem [{\citenamefont {Nishida}(2007)}]{Nishida_2007}%
  \BibitemOpen
  \bibfield  {author} {\bibinfo {author} {\bibfnamefont {Yusuke}\ \bibnamefont
  {Nishida}},\ }\bibfield  {title} {\enquote {\bibinfo {title} {Unitary {F}ermi
  gas at finite temperature in the $\epsilon$ expansion},}\ }\href {\doibase
  10.1103/PhysRevA.75.063618} {\bibfield  {journal} {\bibinfo  {journal} {Phys.
  Rev. A}\ }\textbf {\bibinfo {volume} {75}},\ \bibinfo {pages} {063618}
  (\bibinfo {year} {2007})}\BibitemShut {NoStop}%
\bibitem [{\citenamefont {Veillette}\ \emph {et~al.}(2007)\citenamefont
  {Veillette}, \citenamefont {Sheehy},\ and\ \citenamefont
  {Radzihovsky}}]{Veillette2006}%
  \BibitemOpen
  \bibfield  {author} {\bibinfo {author} {\bibfnamefont {Martin~Y.}\
  \bibnamefont {Veillette}}, \bibinfo {author} {\bibfnamefont {Daniel~E.}\
  \bibnamefont {Sheehy}}, \ and\ \bibinfo {author} {\bibfnamefont {Leo}\
  \bibnamefont {Radzihovsky}},\ }\bibfield  {title} {\enquote {\bibinfo {title}
  {Large-$n$ expansion for unitary superfluid fermi gases},}\ }\href {\doibase
  10.1103/PhysRevA.75.043614} {\bibfield  {journal} {\bibinfo  {journal} {Phys.
  Rev. A}\ }\textbf {\bibinfo {volume} {75}},\ \bibinfo {pages} {043614}
  (\bibinfo {year} {2007})}\BibitemShut {NoStop}%
\bibitem [{\citenamefont {Diehl}\ \emph {et~al.}(2007)\citenamefont {Diehl},
  \citenamefont {Gies}, \citenamefont {Pawlowski},\ and\ \citenamefont
  {Wetterich}}]{Diehl2007}%
  \BibitemOpen
  \bibfield  {author} {\bibinfo {author} {\bibfnamefont {S.}~\bibnamefont
  {Diehl}}, \bibinfo {author} {\bibfnamefont {H.}~\bibnamefont {Gies}},
  \bibinfo {author} {\bibfnamefont {J.~M.}\ \bibnamefont {Pawlowski}}, \ and\
  \bibinfo {author} {\bibfnamefont {C.}~\bibnamefont {Wetterich}},\ }\bibfield
  {title} {\enquote {\bibinfo {title} {Flow equations for the bcs-bec
  crossover},}\ }\href {\doibase 10.1103/PhysRevA.76.021602} {\bibfield
  {journal} {\bibinfo  {journal} {Phys. Rev. A}\ }\textbf {\bibinfo {volume}
  {76}},\ \bibinfo {pages} {021602} (\bibinfo {year} {2007})}\BibitemShut
  {NoStop}%
\bibitem [{\citenamefont {Boettcher}\ \emph {et~al.}(2014)\citenamefont
  {Boettcher}, \citenamefont {Pawlowski},\ and\ \citenamefont
  {Wetterich}}]{Boettcher2014}%
  \BibitemOpen
  \bibfield  {author} {\bibinfo {author} {\bibfnamefont {Igor}\ \bibnamefont
  {Boettcher}}, \bibinfo {author} {\bibfnamefont {Jan~M.}\ \bibnamefont
  {Pawlowski}}, \ and\ \bibinfo {author} {\bibfnamefont {Christof}\
  \bibnamefont {Wetterich}},\ }\bibfield  {title} {\enquote {\bibinfo {title}
  {Critical temperature and superfluid gap of the unitary fermi gas from
  functional renormalization},}\ }\href {\doibase 10.1103/PhysRevA.89.053630}
  {\bibfield  {journal} {\bibinfo  {journal} {Phys. Rev. A}\ }\textbf {\bibinfo
  {volume} {89}},\ \bibinfo {pages} {053630} (\bibinfo {year}
  {2014})}\BibitemShut {NoStop}%
\bibitem [{\citenamefont {Petrov}\ \emph {et~al.}(2004)\citenamefont {Petrov},
  \citenamefont {Salomon},\ and\ \citenamefont {Shlyapnikov}}]{Petrov2004}%
  \BibitemOpen
  \bibfield  {author} {\bibinfo {author} {\bibfnamefont {D.~S.}\ \bibnamefont
  {Petrov}}, \bibinfo {author} {\bibfnamefont {C.}~\bibnamefont {Salomon}}, \
  and\ \bibinfo {author} {\bibfnamefont {G.~V.}\ \bibnamefont {Shlyapnikov}},\
  }\bibfield  {title} {\enquote {\bibinfo {title} {Weakly bound dimers of
  fermionic atoms},}\ }\href {\doibase 10.1103/PhysRevLett.93.090404}
  {\bibfield  {journal} {\bibinfo  {journal} {Phys. Rev. Lett.}\ }\textbf
  {\bibinfo {volume} {93}},\ \bibinfo {pages} {090404} (\bibinfo {year}
  {2004})}\BibitemShut {NoStop}%
\bibitem [{\citenamefont {Salasnich}\ and\ \citenamefont
  {Bighin}(2015)}]{Salasnich2015}%
  \BibitemOpen
  \bibfield  {author} {\bibinfo {author} {\bibfnamefont {L.}~\bibnamefont
  {Salasnich}}\ and\ \bibinfo {author} {\bibfnamefont {G.}~\bibnamefont
  {Bighin}},\ }\bibfield  {title} {\enquote {\bibinfo {title} {Scattering
  length of composite bosons in the three-dimensional bcs-bec crossover},}\
  }\href {\doibase 10.1103/PhysRevA.91.033610} {\bibfield  {journal} {\bibinfo
  {journal} {Phys. Rev. A}\ }\textbf {\bibinfo {volume} {91}},\ \bibinfo
  {pages} {033610} (\bibinfo {year} {2015})}\BibitemShut {NoStop}%
\bibitem [{\citenamefont {Liu}\ \emph {et~al.}(2009)\citenamefont {Liu},
  \citenamefont {Hu},\ and\ \citenamefont {Drummond}}]{Liu2009}%
  \BibitemOpen
  \bibfield  {author} {\bibinfo {author} {\bibfnamefont {Xia-Ji}\ \bibnamefont
  {Liu}}, \bibinfo {author} {\bibfnamefont {Hui}\ \bibnamefont {Hu}}, \ and\
  \bibinfo {author} {\bibfnamefont {Peter~D.}\ \bibnamefont {Drummond}},\
  }\bibfield  {title} {\enquote {\bibinfo {title} {Virial {E}xpansion for a
  {S}trongly {C}orrelated {F}ermi {G}as},}\ }\href
  {http://link.aps.org/doi/10.1103/PhysRevLett.102.160401} {\bibfield
  {journal} {\bibinfo  {journal} {Phys. Rev. Lett.}\ }\textbf {\bibinfo
  {volume} {102}},\ \bibinfo {pages} {160401--} (\bibinfo {year}
  {2009})}\BibitemShut {NoStop}%
\bibitem [{\citenamefont {Li}\ \emph {et~al.}(2021)\citenamefont {Li},
  \citenamefont {Luo}, \citenamefont {Wang}, \citenamefont {Xie}, \citenamefont
  {Liu}, \citenamefont {Hu}, \citenamefont {Chen}, \citenamefont {Yao},\ and\
  \citenamefont {Pan}}]{USTC2021}%
  \BibitemOpen
  \bibfield  {author} {\bibinfo {author} {\bibfnamefont {Xi}~\bibnamefont
  {Li}}, \bibinfo {author} {\bibfnamefont {Xiang}\ \bibnamefont {Luo}},
  \bibinfo {author} {\bibfnamefont {Shuai}\ \bibnamefont {Wang}}, \bibinfo
  {author} {\bibfnamefont {Ke}~\bibnamefont {Xie}}, \bibinfo {author}
  {\bibfnamefont {Xiang-Pei}\ \bibnamefont {Liu}}, \bibinfo {author}
  {\bibfnamefont {Hui}\ \bibnamefont {Hu}}, \bibinfo {author} {\bibfnamefont
  {Yu-Ao}\ \bibnamefont {Chen}}, \bibinfo {author} {\bibfnamefont {Xing-Can}\
  \bibnamefont {Yao}}, \ and\ \bibinfo {author} {\bibfnamefont {Jian-Wei}\
  \bibnamefont {Pan}},\ }\bibfield  {title} {\enquote {\bibinfo {title} {Second
  sound attenuation near quantum criticality},}\ }\href@noop {} {\bibfield
  {journal} {\bibinfo  {journal} {to be published}\ } (\bibinfo {year}
  {2021})}\BibitemShut {NoStop}%
\bibitem [{\citenamefont {Hu}\ \emph {et~al.}(2010)\citenamefont {Hu},
  \citenamefont {Liu},\ and\ \citenamefont {Drummond}}]{Hu2010}%
  \BibitemOpen
  \bibfield  {author} {\bibinfo {author} {\bibfnamefont {Hui}\ \bibnamefont
  {Hu}}, \bibinfo {author} {\bibfnamefont {Xia-Ji}\ \bibnamefont {Liu}}, \ and\
  \bibinfo {author} {\bibfnamefont {Peter~D}\ \bibnamefont {Drummond}},\
  }\bibfield  {title} {\enquote {\bibinfo {title} {Universal thermodynamics of
  a strongly interacting {F}ermi gas: theory versus experiment},}\ }\href
  {http://stacks.iop.org/1367-2630/12/i=6/a=063038} {\bibfield  {journal}
  {\bibinfo  {journal} {New Journal of Physics}\ }\textbf {\bibinfo {volume}
  {12}},\ \bibinfo {pages} {063038} (\bibinfo {year} {2010})}\BibitemShut
  {NoStop}%
\bibitem [{\citenamefont {Tempere}\ and\ \citenamefont
  {Devreese}(2012)}]{tempere2012path}%
  \BibitemOpen
  \bibfield  {author} {\bibinfo {author} {\bibfnamefont {Jacques}\ \bibnamefont
  {Tempere}}\ and\ \bibinfo {author} {\bibfnamefont {Jeroen~PA}\ \bibnamefont
  {Devreese}},\ }\href@noop {} {\emph {\bibinfo {title} {Path-integral
  description of {C}ooper pairing}}}\ (\bibinfo  {publisher} {INTECH Open
  Access Publisher},\ \bibinfo {year} {2012})\BibitemShut {NoStop}%
\bibitem [{\citenamefont {Schirotzek}\ \emph {et~al.}(2008)\citenamefont
  {Schirotzek}, \citenamefont {Shin}, \citenamefont {Schunck},\ and\
  \citenamefont {Ketterle}}]{Schirotzek2008}%
  \BibitemOpen
  \bibfield  {author} {\bibinfo {author} {\bibfnamefont {Andr\'e}\ \bibnamefont
  {Schirotzek}}, \bibinfo {author} {\bibfnamefont {Yong-il}\ \bibnamefont
  {Shin}}, \bibinfo {author} {\bibfnamefont {Christian~H.}\ \bibnamefont
  {Schunck}}, \ and\ \bibinfo {author} {\bibfnamefont {Wolfgang}\ \bibnamefont
  {Ketterle}},\ }\bibfield  {title} {\enquote {\bibinfo {title} {Determination
  of the superfluid gap in atomic {F}ermi gases by quasiparticle
  spectroscopy},}\ }\href {\doibase 10.1103/PhysRevLett.101.140403} {\bibfield
  {journal} {\bibinfo  {journal} {Phys. Rev. Lett.}\ }\textbf {\bibinfo
  {volume} {101}},\ \bibinfo {pages} {140403} (\bibinfo {year}
  {2008})}\BibitemShut {NoStop}%
\bibitem [{\citenamefont {Hoinka}\ \emph {et~al.}(2017)\citenamefont {Hoinka},
  \citenamefont {Dyke}, \citenamefont {Lingham}, \citenamefont {Kinnunen},
  \citenamefont {Bruun},\ and\ \citenamefont {Vale}}]{Hoinka2017}%
  \BibitemOpen
  \bibfield  {author} {\bibinfo {author} {\bibfnamefont {Sascha}\ \bibnamefont
  {Hoinka}}, \bibinfo {author} {\bibfnamefont {Paul}\ \bibnamefont {Dyke}},
  \bibinfo {author} {\bibfnamefont {Marcus~G.}\ \bibnamefont {Lingham}},
  \bibinfo {author} {\bibfnamefont {Jami~J.}\ \bibnamefont {Kinnunen}},
  \bibinfo {author} {\bibfnamefont {Georg~M.}\ \bibnamefont {Bruun}}, \ and\
  \bibinfo {author} {\bibfnamefont {Chris~J.}\ \bibnamefont {Vale}},\
  }\bibfield  {title} {\enquote {\bibinfo {title} {Goldstone mode and
  pair-breaking excitations in atomic {F}ermi superfluids},}\ }\href
  {http://dx.doi.org/10.1038/nphys4187} {\bibfield  {journal} {\bibinfo
  {journal} {Nature Physics}\ }\textbf {\bibinfo {volume} {13}},\ \bibinfo
  {pages} {943} (\bibinfo {year} {2017})}\BibitemShut {NoStop}%
\bibitem [{\citenamefont {Haussmann}\ \emph {et~al.}(2009)\citenamefont
  {Haussmann}, \citenamefont {Punk},\ and\ \citenamefont
  {Zwerger}}]{haussmann2009}%
  \BibitemOpen
  \bibfield  {author} {\bibinfo {author} {\bibfnamefont {R.}~\bibnamefont
  {Haussmann}}, \bibinfo {author} {\bibfnamefont {M.}~\bibnamefont {Punk}}, \
  and\ \bibinfo {author} {\bibfnamefont {W.}~\bibnamefont {Zwerger}},\
  }\bibfield  {title} {\enquote {\bibinfo {title} {Spectral functions and rf
  response of ultracold fermionic atoms},}\ }\href {\doibase
  10.1103/PhysRevA.80.063612} {\bibfield  {journal} {\bibinfo  {journal} {Phys.
  Rev. A}\ }\textbf {\bibinfo {volume} {80}},\ \bibinfo {pages} {063612}
  (\bibinfo {year} {2009})}\BibitemShut {NoStop}%
\bibitem [{\citenamefont {Hu}\ \emph {et~al.}(2011)\citenamefont {Hu},
  \citenamefont {Liu},\ and\ \citenamefont {Drummond}}]{Hu2011b}%
  \BibitemOpen
  \bibfield  {author} {\bibinfo {author} {\bibfnamefont {Hui}\ \bibnamefont
  {Hu}}, \bibinfo {author} {\bibfnamefont {Xia-Ji}\ \bibnamefont {Liu}}, \ and\
  \bibinfo {author} {\bibfnamefont {Peter~D}\ \bibnamefont {Drummond}},\
  }\bibfield  {title} {\enquote {\bibinfo {title} {Universal contact of
  strongly interacting fermions at finite temperatures},}\ }\href {\doibase
  10.1088/1367-2630/13/3/035007} {\bibfield  {journal} {\bibinfo  {journal}
  {New Journal of Physics}\ }\textbf {\bibinfo {volume} {13}},\ \bibinfo
  {pages} {035007} (\bibinfo {year} {2011})}\BibitemShut {NoStop}%
\bibitem [{\citenamefont {Sagi}\ \emph {et~al.}(2015)\citenamefont {Sagi},
  \citenamefont {Drake}, \citenamefont {Paudel}, \citenamefont {Chapurin},\
  and\ \citenamefont {Jin}}]{Sagi2014}%
  \BibitemOpen
  \bibfield  {author} {\bibinfo {author} {\bibfnamefont {Yoav}\ \bibnamefont
  {Sagi}}, \bibinfo {author} {\bibfnamefont {Tara~E.}\ \bibnamefont {Drake}},
  \bibinfo {author} {\bibfnamefont {Rabin}\ \bibnamefont {Paudel}}, \bibinfo
  {author} {\bibfnamefont {Roman}\ \bibnamefont {Chapurin}}, \ and\ \bibinfo
  {author} {\bibfnamefont {Deborah~S.}\ \bibnamefont {Jin}},\ }\bibfield
  {title} {\enquote {\bibinfo {title} {Breakdown of the {F}ermi liquid
  description for strongly interacting fermions},}\ }\href {\doibase
  10.1103/PhysRevLett.114.075301} {\bibfield  {journal} {\bibinfo  {journal}
  {Phys. Rev. Lett.}\ }\textbf {\bibinfo {volume} {114}},\ \bibinfo {pages}
  {075301} (\bibinfo {year} {2015})}\BibitemShut {NoStop}%
\bibitem [{\citenamefont {Laurent}\ \emph {et~al.}(2017)\citenamefont
  {Laurent}, \citenamefont {Pierce}, \citenamefont {Delehaye}, \citenamefont
  {Yefsah}, \citenamefont {Chevy},\ and\ \citenamefont
  {Salomon}}]{Laurent2017}%
  \BibitemOpen
  \bibfield  {author} {\bibinfo {author} {\bibfnamefont {S\'ebastien}\
  \bibnamefont {Laurent}}, \bibinfo {author} {\bibfnamefont {Matthieu}\
  \bibnamefont {Pierce}}, \bibinfo {author} {\bibfnamefont {Marion}\
  \bibnamefont {Delehaye}}, \bibinfo {author} {\bibfnamefont {Tarik}\
  \bibnamefont {Yefsah}}, \bibinfo {author} {\bibfnamefont {Fr\'ed\'eric}\
  \bibnamefont {Chevy}}, \ and\ \bibinfo {author} {\bibfnamefont {Christophe}\
  \bibnamefont {Salomon}},\ }\bibfield  {title} {\enquote {\bibinfo {title}
  {Connecting few-body inelastic decay to quantum correlations in a many-body
  system: A weakly coupled impurity in a resonant {F}ermi gas},}\ }\href
  {\doibase 10.1103/PhysRevLett.118.103403} {\bibfield  {journal} {\bibinfo
  {journal} {Phys. Rev. Lett.}\ }\textbf {\bibinfo {volume} {118}},\ \bibinfo
  {pages} {103403} (\bibinfo {year} {2017})}\BibitemShut {NoStop}%
\bibitem [{\citenamefont {Mukherjee}\ \emph {et~al.}(2019)\citenamefont
  {Mukherjee}, \citenamefont {Patel}, \citenamefont {Yan}, \citenamefont
  {Fletcher}, \citenamefont {Struck},\ and\ \citenamefont
  {Zwierlein}}]{Mukherjee2019}%
  \BibitemOpen
  \bibfield  {author} {\bibinfo {author} {\bibfnamefont {Biswaroop}\
  \bibnamefont {Mukherjee}}, \bibinfo {author} {\bibfnamefont {Parth~B.}\
  \bibnamefont {Patel}}, \bibinfo {author} {\bibfnamefont {Zhenjie}\
  \bibnamefont {Yan}}, \bibinfo {author} {\bibfnamefont {Richard~J.}\
  \bibnamefont {Fletcher}}, \bibinfo {author} {\bibfnamefont {Julian}\
  \bibnamefont {Struck}}, \ and\ \bibinfo {author} {\bibfnamefont {Martin~W.}\
  \bibnamefont {Zwierlein}},\ }\bibfield  {title} {\enquote {\bibinfo {title}
  {Spectral response and contact of the unitary {F}ermi gas},}\ }\href
  {\doibase 10.1103/PhysRevLett.122.203402} {\bibfield  {journal} {\bibinfo
  {journal} {Phys. Rev. Lett.}\ }\textbf {\bibinfo {volume} {122}},\ \bibinfo
  {pages} {203402} (\bibinfo {year} {2019})}\BibitemShut {NoStop}%
\bibitem [{\citenamefont {Carcy}\ \emph {et~al.}(2019)\citenamefont {Carcy},
  \citenamefont {Hoinka}, \citenamefont {Lingham}, \citenamefont {Dyke},
  \citenamefont {Kuhn}, \citenamefont {Hu},\ and\ \citenamefont
  {Vale}}]{Carcy2019}%
  \BibitemOpen
  \bibfield  {author} {\bibinfo {author} {\bibfnamefont {C.}~\bibnamefont
  {Carcy}}, \bibinfo {author} {\bibfnamefont {S.}~\bibnamefont {Hoinka}},
  \bibinfo {author} {\bibfnamefont {M.~G.}\ \bibnamefont {Lingham}}, \bibinfo
  {author} {\bibfnamefont {P.}~\bibnamefont {Dyke}}, \bibinfo {author}
  {\bibfnamefont {C.~C.~N.}\ \bibnamefont {Kuhn}}, \bibinfo {author}
  {\bibfnamefont {H.}~\bibnamefont {Hu}}, \ and\ \bibinfo {author}
  {\bibfnamefont {C.~J.}\ \bibnamefont {Vale}},\ }\bibfield  {title} {\enquote
  {\bibinfo {title} {Contact and sum rules in a near-uniform {F}ermi gas at
  unitarity},}\ }\href {\doibase 10.1103/PhysRevLett.122.203401} {\bibfield
  {journal} {\bibinfo  {journal} {Phys. Rev. Lett.}\ }\textbf {\bibinfo
  {volume} {122}},\ \bibinfo {pages} {203401} (\bibinfo {year}
  {2019})}\BibitemShut {NoStop}%
\bibitem [{\citenamefont {Tempere}\ \emph {et~al.}(2008)\citenamefont
  {Tempere}, \citenamefont {Klimin}, \citenamefont {Devreese},\ and\
  \citenamefont {Moshchalkov}}]{tempere_d_wave}%
  \BibitemOpen
  \bibfield  {author} {\bibinfo {author} {\bibfnamefont {J.}~\bibnamefont
  {Tempere}}, \bibinfo {author} {\bibfnamefont {S.~N.}\ \bibnamefont {Klimin}},
  \bibinfo {author} {\bibfnamefont {J.~T.}\ \bibnamefont {Devreese}}, \ and\
  \bibinfo {author} {\bibfnamefont {V.~V.}\ \bibnamefont {Moshchalkov}},\
  }\bibfield  {title} {\enquote {\bibinfo {title} {Imbalanced $d$-wave
  superfluids in the {BCS-BEC} crossover regime at finite temperatures},}\
  }\href {\doibase 10.1103/PhysRevB.77.134502} {\bibfield  {journal} {\bibinfo
  {journal} {Phys. Rev. B}\ }\textbf {\bibinfo {volume} {77}},\ \bibinfo
  {pages} {134502} (\bibinfo {year} {2008})}\BibitemShut {NoStop}%
\bibitem [{\citenamefont {Nussinov}\ and\ \citenamefont
  {Nussinov}(2006)}]{Nussinov2006}%
  \BibitemOpen
  \bibfield  {author} {\bibinfo {author} {\bibfnamefont {Zohar}\ \bibnamefont
  {Nussinov}}\ and\ \bibinfo {author} {\bibfnamefont {Shmuel}\ \bibnamefont
  {Nussinov}},\ }\bibfield  {title} {\enquote {\bibinfo {title} {Triviality of
  the bcs-bec crossover in extended dimensions: Implications for the ground
  state energy},}\ }\href {\doibase 10.1103/PhysRevA.74.053622} {\bibfield
  {journal} {\bibinfo  {journal} {Phys. Rev. A}\ }\textbf {\bibinfo {volume}
  {74}},\ \bibinfo {pages} {053622} (\bibinfo {year} {2006})}\BibitemShut
  {NoStop}%
\end{thebibliography}%
\end{document}